\documentclass[journal,article,submit,pdftex,moreauthors]{Definitions/mdpi}
\usepackage{listings}
\usepackage{siunitx}
\newcommand{\farcs}{\mbox{$.\!\!^{\prime\prime}$}}

\firstpage{1} 
\pubvolume{1}
\issuenum{1}
\articlenumber{0}
\pubyear{2025}
\copyrightyear{2025}
\datereceived{ } 
\daterevised{ } 
\dateaccepted{ } 
\datepublished{ } 
\hreflink{https://doi.org/} 

\Title{From Non-parametric Statistics to Images: Generating Galaxy Morphologies}

\TitleCitation{From Non-parametric Statistics to Images: Generating Galaxy Morphologies}

\Author{Quanfeng Xu $^{1,2}$\orcidA{}, 
Mi Chen $^{5}$\orcidB{}, 
Renhao Ye $^{1,2}$\orcidC{}, 
Siyuan Chen $^{1,2}$\orcidF{}, 
Nan Li $^{4}$\orcidD{} and Shiyin Shen $^{2,3,*}$\orcidE{}}

\AuthorNames{Quanfeng Xu, Mi Chen, Renhao Ye, Siyuan Chen, Nan Li and Shiyin Shen}

\isAPAStyle{%
       \AuthorCitation{Xu, Q., Chen, M., Ye, R., Chen, S., Li, N., \& Shen, S.}
         }{%
        \isChicagoStyle{%
        \AuthorCitation{Xu, Quanfeng, Mi Chen, Renhao Ye, Siyuan Chen, Nan Li and Shiyin Shen.}
        }{
        \AuthorCitation{Xu, Q.; Chen, M.; Ye, R.; Chen, S.; Li, N.; Shen, S.}
        }
}

\address{%
$^{1}$ \quad School of Astronomy and Space Science, University of Chinese Academy of Sciences, 1 East Yanqi Lake Rd., Beijing 100049, China\\
$^{2}$ \quad Shanghai Astronomical Observatory, Chinese Academy of Sciences, 80 Nandan Rd., Shanghai 200030, China\\
$^{3}$ \quad Key Lab for Astrophysics, Shanghai 200234, China\\
$^{4}$ \quad National Astronomical Observatory, Chinese Academy of Sciences, Beijing 100101, China\\
$^{5}$ \quad Kapteyn Astronomical Institute, University of Groningen, Postbus 800, 9700 AV Groningen, The Netherlands}

\corres{Correspondence: ssy@shao.ac.cn (S.S.)}

\abstract{Non-parametric statistics characterize complex galaxy morphologies but do not specify how to construct images that reproduce those statistics. We present Flow-VAE, which combines a variational autoencoder with a conditional normalizing flow to generate images conditioned on concentration, asymmetry, smoothness ($CAS$), Gini coefficient ($G$), and $M_{20}$. A Gaussian mixture model completes unspecified descriptors and measures the training-distribution support of requested combinations. Using a dataset of 240,000 DESI Legacy Imaging Surveys $r$-band images, we evaluate fidelity by remeasuring morphological statistics from model outputs and comparing them with the inputs. With a tolerance of $\pm10\%$ of each descriptor’s adopted range, descriptor-specific recovery rates are 96.62–99.97\% for reconstruction and 87.81–99.83\% for single-parameter generation among valid measurements. 
Generated descriptors show regression toward typical training values, while less-supported joint conditions produce broader residuals and systematic offsets. These results quantify the extent to which Flow-VAE can generate galaxy images consistent with prescribed non-parametric morphological statistics.}


\nolinenumbers
\renewcommand{\linenumbers}[1][]{}
\begin{document}
\nolinenumbers

\section{Introduction}\label{sec:introduction}

Galaxies are intrinsically complex three-dimensional structures that are projected into two-dimensional morphological images in observational astronomy. Analyzing these morphologies provides essential insights into galaxy formation and evolution \citep{Robertson2019GalaxyFA}. Traditionally, visual classification schemes such as the Hubble sequence \citep{1926ApJ....64..321H} were used to categorize these structures; however, such subjective approaches are often too simplistic to capture their full structural diversity and evolutionary intricacies. By mapping the distribution of surface brightness \citep{10.1126/science.1060855}, features such as spiral arms, bulges, and disks reveal underlying physical processes, including mergers, star formation, and internal evolution. This morphological information helps determine the mass distribution of a galaxy, its interactions with the environment, and the evolution of large-scale cosmic structures, serving as a vital observational test for cosmological models.

A widely used approach to quantitative morphology is to represent the galaxy surface-brightness distribution with an analytical profile. For example, the S{\'e}rsic profile describes the radial light distribution using the S{\'e}rsic index ($n$), with lower and higher values generally corresponding to disk-dominated and centrally concentrated systems, respectively. Many profile-fitting tools have been developed to efficiently fit galaxy surface-brightness distributions, including \texttt{GALFIT} \citep{Peng_2010}, \texttt{ProFit} \citep{10.1093/mnras/stw3039}, and \texttt{GalMOSS} \citep{CHEN2024100825}. Although these models provide compact and interpretable descriptions of the global galaxy structure, their results depend on the adopted profile, initialization, and fitting procedure. Moreover, a small number of smooth analytical components cannot always reproduce irregular substructures, asymmetries, or clumpy emission, particularly for high-redshift and interacting galaxies, potentially biasing the inferred morphological properties \citep{10.1111/j.1365-2966.2010.17690.x}.


Non-parametric statistics provide a complementary description of galaxy morphology by measuring the pixel-level flux distribution without assuming an explicit analytical surface-brightness profile. One of the most widely used systems is Concentration, Asymmetry, and Smoothness (CAS) \citep{Conselice_2003}, where concentration measures the central concentration of light, asymmetry quantifies deviations from rotational symmetry, and smoothness characterizes small-scale structures. Other commonly used descriptors include the Gini coefficient, $M_{20}$, and the multimode, intensity, and deviation statistics \citep[$MID$,][]{10.1093/mnras/stt1016}, which capture complementary aspects of flux inequality, the spatial distribution of bright regions, and multi-component structure. These measurements can be computed efficiently using tools such as \texttt{statmorph} \citep{10.1093/mnras/sty3345} and its CSST-oriented extension \texttt{statmorph\_csst}\footnote{\texttt{statmorph\_csst}: \url{https://github.com/pentyum/statmorph_csst.git}.}. Non-parametric statistics have consequently been widely applied to studies of galaxy mergers \citep{Conselice_2003}, star formation \citep{Wuyts_2011}, and morphological evolution \citep{10.1111/j.1365-2966.2008.13936.x}, as well as to the study of the morphological evolution of the galaxy population as a whole \citep{Conselice2014}.

Two properties make these indices particularly well suited to the present problem. First, they are model-independent: they are computed from the pixel distribution of the galaxy and require no assumed surface-brightness profile, so they remain well-defined for clumpy, asymmetric, and interacting systems for which a S{\'e}rsic fit is ill-conditioned. Second, because they are defined by a measurement recipe rather than by a fitted model, the same estimator can be applied without modification to observed images, to mock images drawn from cosmological simulations \citep{10.1093/mnras/sty3345}, and to synthetic images produced by a generative model. This last property is what makes the closed-loop validation performed in this work possible.

Because these five descriptors are strongly correlated in the observed galaxy population, specifying one of them does not specify the others, and sampling the remaining four independently would produce combinations that are rare or absent in the data. We therefore model the joint distribution of $(C,A,S,G,M_{20})$ with a Gaussian mixture model (GMM), a weighted sum of $K$ multivariate Gaussians whose weights, means, and covariances are fitted to the training sample. A GMM provides a flexible density estimate in a low-dimensional space, and we can evaluate both its conditional distributions and its density in closed form. This is what allows a partially specified condition vector to be completed in a way that respects the observed correlations (Section~\ref{sec:gmm_prior}), and it supplies an explicit and interpretable measure of how well a requested combination is represented in the training data (Section~\ref{sec:sample_ood}).

However, the absence of an explicit generative profile makes the inverse problem fundamentally underdetermined. Non-parametric statistics compress a two-dimensional surface-brightness distribution into a small set of scalar quantities, and morphologically distinct galaxies can therefore share similar values of $C$, $A$, $S$, $G$, and $M_{20}$. Although these statistics can be measured directly from an observed image, no unique analytical mapping exists from a prescribed set of descriptors to a realistic galaxy image. This limitation becomes particularly important when the goal is not only to characterize observed galaxies but also to generate simulated galaxies with controlled morphological properties.

Traditional simulation tools such as \texttt{GalSim} \citep{2015AandC....10..121R} efficiently generate galaxy images from analytical surface-brightness profiles. However, these profiles do not naturally reproduce or directly control the asymmetric, clumpy, and multi-component structures captured by non-parametric statistics. Deep generative models offer a potential solution by learning the distribution of galaxy images directly from observational data, thereby generating realistic morphologies beyond simple analytical profiles. GalSim-Hub \citep{Lanusse2021}, for example, developed a Flow-VAE conditioned on galaxy size, magnitude, and photometric redshift. 
For Euclid image simulations, \citet{Euclid} conditioned a generative model on the half-light radius, axis ratio, and S{\'e}rsic index. Compared with conventional parametric simulations, these data-driven approaches can reproduce a richer range of galaxy morphologies and more realistic substructures. Table~\ref{tab:conditioning} summarizes how their conditioning variables compare with those adopted here.

\begin{table*}[t]
\caption{Comparison of the conditioning variables used by data-driven galaxy image simulation and generation methods.}
\label{tab:conditioning}
\centering
\small
\begin{tabularx}{\textwidth}{l l l}
\toprule
Study & Conditioning variables & Type of variable \\
\midrule
\texttt{GalSim} \citep{2015AandC....10..121R} & Profile parameters & Parametric \\
GalSim-Hub \citep{Lanusse2021} & Magnitude, size, $z_{\rm phot}$ & Global photometric \\
Euclid \citep{Euclid} & Size, magnitude, $z$, axis ratio, S{\'e}rsic $n$ & Mixed parametric \\
This work & $C$, $A$, $S$, $G$, $M_{20}$ & Non-parametric, measured \\
\bottomrule
\end{tabularx}
\end{table*}

Nevertheless, the conditioning variables employed by these two generative models predominantly encode global properties of galaxy structure, and thus provide limited control over higher-order morphological features, such as asymmetry, smoothness, and clumpiness. In addition, parameters such as the half-light radius and the S{\'e}rsic index are sensitive to the specific measurement protocol or profile-fitting methodology adopted, complicating the assessment of whether a stochastically generated image faithfully satisfies the prescribed constraints.  Therefore, Conditioning generation on directly measurable, non-parametric summary statistics would not only be beneficial for simulation studies that require detailed morphological features of galaxies, such as the modeling of the morphological characteristics of sources in strong gravitational lensing  \citep{10.1093/mnras/stae1577},  but also allow quantitative checks that generated galaxies match requested properties.

In fact, for stochastic generative models, visual realism and agreement with the population distribution do not ensure that individual outputs satisfy their prescribed conditions. Remeasuring the conditioning statistics from generated images and comparing them directly with the inputs therefore provides an essential object-level validation. In this study, we develop a Flow-VAE model conditioned on five widely used non-parametric morphological statistics: concentration ($C$), asymmetry ($A$), smoothness ($S$), the Gini coefficient ($G$), and $M_{20}$. Our objective is to quantify conditional fidelity by comparing the requested morphological statistics with those remeasured from generated images. This closed-loop evaluation is a key methodological contribution of our framework, as it quantifies conditional accuracy and reveals systematic biases across the sampled parameter space.


The remainder of this paper is organized as follows. Section~\ref{sec:data} describes the galaxy sample, image preprocessing, and measurement of the non-parametric statistics. Section~\ref{sec:methods} introduces the Flow-VAE architecture and the GMM used to model the joint descriptor distribution. Section~\ref{sec:results} evaluates reconstruction fidelity, single-parameter control, and generation under low-density conditions, and validates the generated images against structural diagnostics that are not used as conditioning variables. Section~\ref{sec:discussion} discusses the model's capabilities and limitations, and Section~\ref{sec:conclusions} summarizes the main conclusions.

\section{Datasets}\label{sec:data}


\subsection{Galaxy Images and Preprocessing}\label{sec:images}

We construct a galaxy sample from the SDSS main galaxy sample (MGS; $r_{\rm petro}<17.77$) \citep{Ahn2012}, using imaging from the DESI Legacy Imaging Surveys \citep{Dey_2019}. Table~\ref{tab:cutflow} summarizes the selection procedure and the number of objects remaining at each step. For model training, we use $64\times64$ $r$-band cutouts with a pixel scale of $0\farcs262$. This cutout size was chosen to keep the computational cost manageable.

Following the catalogue matching and source-detection requirements listed in Table~\ref{tab:cutflow}, we apply quality and range cuts to the 42 parameters measured by SExtractor and \texttt{statmorph}. The final selection retains galaxies with $r\gtrsim15$ and $z\leq0.3$, excludes the most extended and most compact sources, and requires sufficient space within each cutout for both the galaxy measurement aperture and a background annulus. We draw a working sample of \num{240000} galaxies from the surviving objects. These requirements favour compact galaxies of intermediate surface brightness at fixed magnitude, so the model describes this selected population rather than the full MGS.

\begin{table}[H]
\caption{Sample selection and the number of objects remaining after each step. The final row combines the quality and range cuts with the selection of the working sample.}
\label{tab:cutflow}
\centering
\small
\begin{tabularx}{\columnwidth}{c X r}
\toprule
Step & Selection criterion & $N$ \\
\midrule
0 & SDSS MGS: $(\mathrm{VAGC\_SELECT}\,\&\,4)\ne0$,
    $0<z<0.5$, and $r_{\rm petro}<17.77$
    & 633,568 \\
1 & DECaLS southern imaging footprint:
    NGC with $\delta<32.375^\circ$, and the full SGC
    & 410,184 \\
2 & Match to a DR9 BGS\_BRIGHT target within $0\farcs79$
    & 394,465 \\
3 & Availability of a $64\times64$ cutout
    & 374,203 \\
4 & SExtractor detection within the central $8\times8$-pixel region in the $g$, $r$, and $z$ bands
    & 371,490 \\
5 & Quality and range cuts, followed by selection
    of the working sample
    & 240,000 \\
\bottomrule
\end{tabularx}
\end{table}

\begin{figure}
    \centering
    \includegraphics[width=\textwidth]{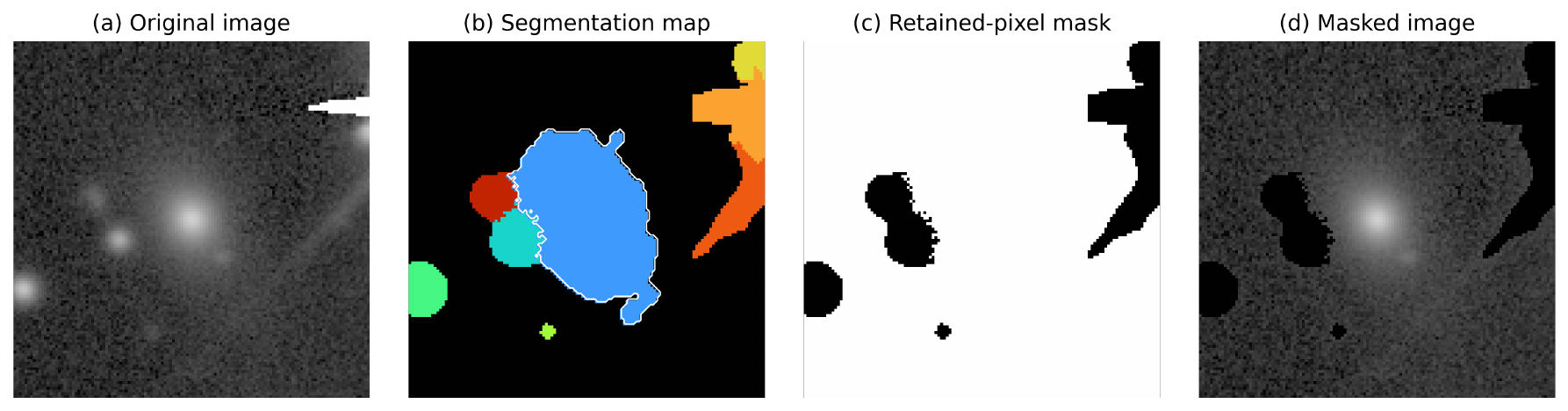}
    \caption{
    Example of contaminant masking.
    (a) Original image.
    (b) SExtractor segmentation map.
    (c) Binary mask retaining the target galaxy and background.
    (d) Masked image.
    The white contour in (b) outlines the target galaxy; black regions in (c) and (d) indicate excluded sources.
    }
    \label{fig:preprocessing}
\end{figure}

Because non-parametric morphological statistics are derived from the image flux distribution, contamination from foreground stars and neighbouring galaxies can bias their measurements. We therefore perform source detection on each cutout using \texttt{SExtractor} \citep{refId0}, adopting a detection threshold of $1.5\sigma$ above the local background and a minimum detection area of five connected pixels. Each cutout is centred on the catalogued position of an MGS galaxy, and the detection containing the central pixel is identified as the target. All other detections are treated as potential contaminants. From the resulting segmentation map, we construct a binary mask that retains pixels assigned to the target galaxy and the background, while excluding pixels assigned to other detected sources. Figure~\ref{fig:preprocessing} illustrates this procedure for a representative cutout. The resulting masks are used in the morphological measurements and the reconstruction loss to reduce contamination from neighbouring sources.




\subsection{Non-parametric Statistics}\label{sec:morph_parameter}

Before performing non-parametric statistical analyses on each galaxy sample, we use the previously generated \texttt{SExtractor} masks to ensure that only pixels belonging to the target galaxy contribute to the calculations. We then utilize \texttt{Statmorph} \citep{10.1093/mnras/sty3345} on the masked images to extract the galaxies' non-parametric morphological statistics, which comprise \textit{concentration} ($C$), \textit{asymmetry} ($A$), \textit{smoothness} ($S$) \citep[$CAS$,][]{Conselice_2003}, the Gini coefficient \citep[$G$,][]{Lotz_2004}, and the second-order moment of the brightest 20\% of the galaxy's flux \citep[$M_{20}$,][]{Lotz_2004}. The definitions of these non-parametric morphological statistics are listed below.

\begin{itemize}

\item \textbf{CAS}

\begin{itemize}

\item \textit{Concentration} ($C$) quantifies how concentrated the light is within a galaxy by comparing the radii that contain 20\% and 80\% of the total light. It is calculated using the formula:
\begin{equation}\label{equ:Concentration}
C = 5 \log_{10} \left( \frac{r_{80}}{r_{20}} \right)
\end{equation}

\item \textit{Asymmetry} ($A$) measures the rotational symmetry of a galaxy by rotating its image 180 degrees and comparing it to the original. The calculation is given by:
\begin{equation}\label{equ:Asymmetry}
A = \frac{\sum |I - I_{180^{\circ}}|}{\sum |I|} - B_A
\end{equation}
where $I$ is the galaxy image and $B_A$ is the dimensionless background-asymmetry correction.

\item \textit{Smoothness} ($S$) quantifies the presence of small-scale structures, such as star clusters, by comparing a smoothed version of the galaxy image to the original. It is calculated as:
\begin{equation}\label{equ:Smoothness}
S = \frac{\sum |I - I_{S}|}{\sum |I|} - B_S
\end{equation}
where $I_{S}$ is the galaxy’s image smoothed by a boxcar of width 0.25 times the petrosian radius $r_p$. The term $B_S$ is a background correction that accounts for small-scale fluctuations introduced by sky noise.

\end{itemize}
\item \textbf{G}: The \textit{Gini coefficient} ($G$) measures the distribution of light across the pixels of a galaxy, with values ranging from 0 (uniform light distribution) to 1 (highly concentrated light). It is defined as:
\begin{equation}\label{equ:Gini_coefficient}
G = \frac{1}{2 \bar{I} n (n - 1)} \sum_{i=1}^{n} \sum_{j=1}^{n} |I_i - I_j|
\end{equation}
where $\bar{I}$ represents the mean brightness of all pixels. A higher Gini coefficient reflects light concentrated in a few regions.
\item \textbf{$M_{20}$}: The $M_{20}$ metric represents the second-order moment of the brightest 20\% of the pixels, capturing the spread of light and highlighting bright off-centre regions. It is calculated as:
\begin{equation}\label{equ:Mi}
M_{\text{tot}} = \sum_i^n M_i = \sum_i^n f_i \cdot ((x_i-x_{\text{cen}})^2+(y_i-y_{\text{cen}})^2)
\end{equation}
\begin{equation}\label{equ:M20}
M_{20} = \log_{10} \left( \frac{\sum_i M_i}{M_{\text{tot}}} \right), \sum_i f_i < 0.2 f_{\text{tot}}
\end{equation}
where $(x_{\text{cen}}, y_{\text{cen}})$ denotes the galaxy’s centroid, and $M_i$ is the product of each bright pixel’s flux $f_i$ and the square of its distance from the galaxy center.
\end{itemize}

\begin{figure*}
	\includegraphics[width=\textwidth]{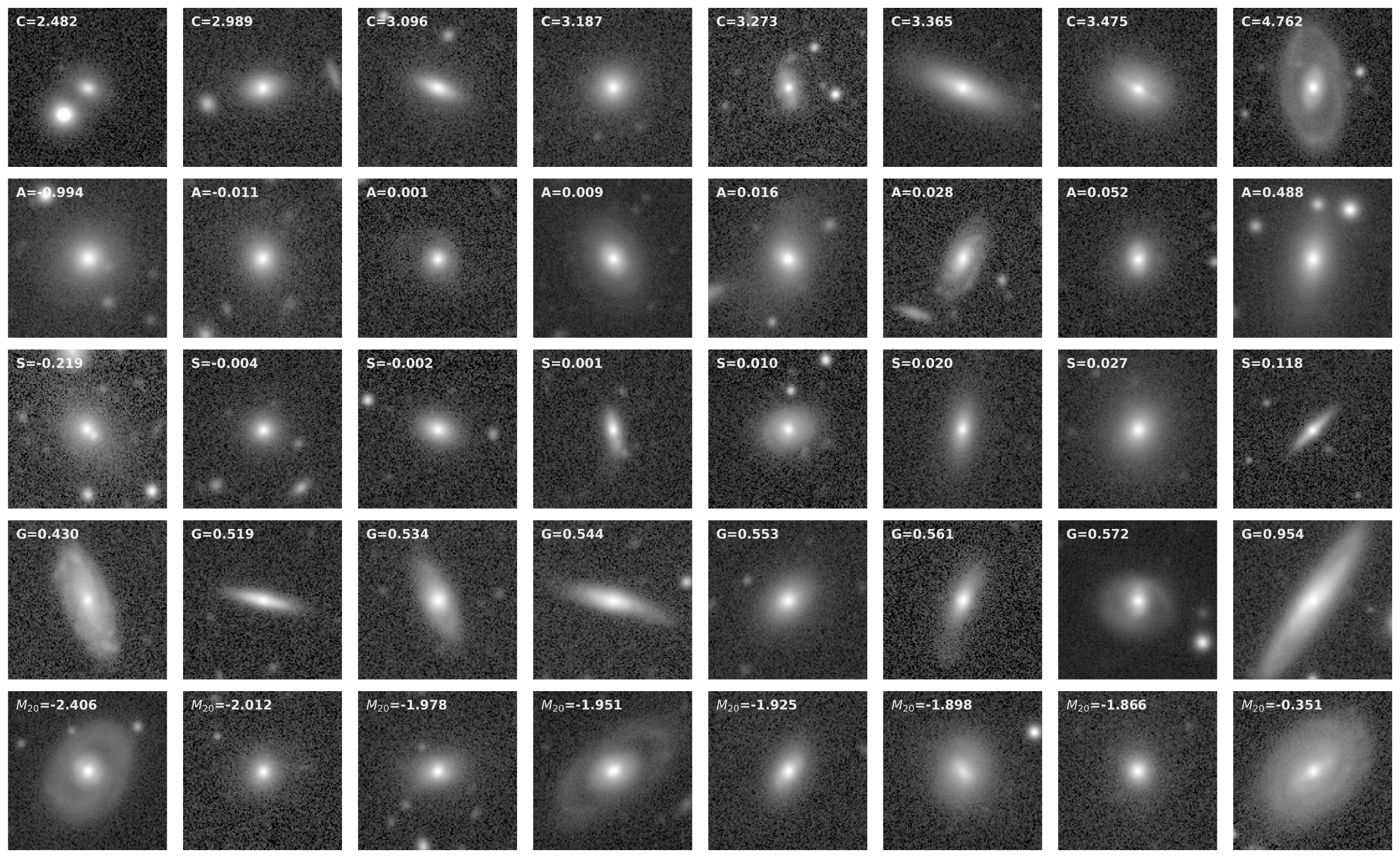}
    \caption{
    Visualizing the physical interpretation of non-parametric morphological statistics through controlled sampling of real galaxy images. Each row isolates the effect of a single target parameter, which is uniformly sampled across its numerical range from left to right, while the other four parameters are constrained near their overall median values. The visual sequences explicitly demonstrate the structural transitions captured by each statistic: $C$ (Diffuse $\rightarrow$ Concentrated), $A$ (Symmetric $\rightarrow$ Asymmetric), $S$ (Smooth $\rightarrow$ Clumpy/Structured), $G$ (Uniform $\rightarrow$ Concentrated Flux), and $M_{20}$ (Core-dominated $\rightarrow$ Extended/Multi-core). The exact value of the varying parameter is annotated in the top-left corner of each panel.
    }
    \label{fig:show}
\end{figure*}

We measured the non-parametric morphological statistics of \num{240000} $r$-band galaxy images using the \texttt{Statmorph} package. Although several of these statistics are formally defined over much broader domains, such as $M_{20} \in (-\infty, 0]$ and $G \in [0, 1]$, the values measured for real galaxies occupy substantially narrower regions of parameter space. To exclude rare extreme measurements and restrict the generative model to morphologies representative of the DESI MGS sample, we adopt ranges that contain approximately 99\% of the observed values: $C \in [0.5,4.5]$, $A \in [-0.7,1.0]$, $S \in [-0.25,0.2]$, $G \in [0.4,0.95]$, and $M_{20} \in [-2.5,-0.5]$.

Among the galaxies analysed $23.14\%$ have $A<0$, and $34.43\%$ have $S<0$. Negative values can arise when the background correction exceeds the normalized image residual, but do not by themselves establish that the corresponding morphological structure is undetectable. We retain these values as conditioning targets because the model aims to reproduce measured, background-corrected descriptors.

Figure~\ref{fig:show} illustrates how each statistic relates to galaxy structure within these adopted ranges. For each row, one target parameter is uniformly sampled across its allowed interval, while the remaining four parameters are constrained near their median values in the dataset. This controlled sampling isolates the morphological variation associated with each statistic and provides a direct visual interpretation of the parameter space used for model training and generation.



\section{Methods}\label{sec:methods}


In this section, we first briefly introduce the principles of the VAE (Section \ref{sec:vae}) models and Flow (Section \ref{sec:flow}). We then use Flow-VAE (Section \ref{sec:framework}) to map the distributions of non-parametric statistics, latent variables, and galaxy images to each other.

\subsection{Variational Auto-Encoders}\label{sec:vae}

Variational Autoencoders \citep[VAEs;][]{kingma2013auto} learn probabilistic latent representations of complex data distributions \citep{10.1093/mnras/stad3181}. Unlike conventional autoencoders, which map each input to a fixed latent representation, a VAE represents each input as a probability distribution in latent space. A VAE consists of three main components: an encoder, a variational inference mechanism, and a decoder. The encoder maps a galaxy image $\mathbf{x}$ to a distribution in an $N$-dimensional latent space, typically parameterized as a Gaussian distribution with mean $\mu(\mathbf{x})$ and variance $\sigma(\mathbf{x})^2$:

\begin{equation}\label{equ:var_infer}
\mathbf{z} \sim q_\phi(\mathbf{z}\mid\mathbf{x}) = 
\mathcal{N}\left(
\mu(\mathbf{x}),
\mathrm{diag}\left[\sigma(\mathbf{x})^2\right]
\right),
\qquad
\mathbf{z}\in\mathbb{R}^{N}.
\end{equation}

Here, $q_\phi(\mathbf{z}\mid\mathbf{x})$ denotes the approximate posterior distribution of the latent variables inferred from the input galaxy image. During training, a latent vector $\mathbf{z}$ is sampled from this posterior and passed to the decoder, which maps it back to image space to produce the reconstructed image $\hat{\mathbf{x}}$. To regularize the latent space, the VAE minimizes the \textit{Kullback–Leibler} divergence \citep[KL;][]{Kullback1951} between the approximate posterior and a predefined prior distribution. The overall loss is therefore

\begin{equation}\label{equ:vae_loss}
\mathcal L_{\rm VAE} = 
\mathcal L_{\rm rec}
+
\lambda_{\rm KL}
D_{\rm KL}\left(
q_\phi(\mathbf z\mid\mathbf x)
||p_{\rm prior}(\mathbf z)
\right),
\end{equation}

where $\mathcal L_{\rm rec}$ measures the reconstruction error between $\mathbf{x}$ and $\hat{\mathbf{x}}$, $p_{\rm prior}(\mathbf z)$ is the adopted latent prior, and $\lambda_{\rm KL}$ controls the balance between reconstruction fidelity and posterior regularization. This formulation encourages a continuous and regularized latent space while preserving sufficient information for image reconstruction and generation.

\subsection{Normalizing Flow}\label{sec:flow}

Normalizing flows model complex probability densities by transforming a simple base distribution through a sequence of invertible and differentiable mappings, $f_{\theta} = f_{\theta_N} \circ \dots \circ f_{\theta_1}$ \citep{DinhKB14, DurkanB0P19}. In the context of our VAE framework, normalizing flows are utilized to learn the conditional latent density $p_\theta(\mathbf{z}\mid\mathbf{y})$ \citep{Papamakarios2017}. Here, $\mathbf{z}$ represents the VAE latent vector, and $\mathbf{y} = (C,A,S,G,M_{20})$ denotes the vector of non-parametric morphological statistics.

For a fixed context $\mathbf{y}$, the forward transformation maps the VAE latent vectors $\mathbf{z}$ to a simple base distribution variable $\mathbf{u}$:
\begin{equation}\label{equ:trans}
\mathbf u=f_\theta(\mathbf z;\mathbf y),
\qquad
\mathbf z=f_\theta^{-1}(\mathbf u;\mathbf y),
\qquad
\mathbf z,\mathbf u\in\mathbb{R}^{N}.
\end{equation}
Crucially, the morphological descriptors $\mathbf{y}$ condition the transformation; they are not its transformed output. Key properties of normalizing flows include differentiability, which ensures the Jacobian determinant can be computed, and invertibility enables the probability density to be evaluated exactly through the change-of-variables formula.

We employ a Masked Autoregressive Flow \citep[MAF,][]{Papamakarios2017} for its expressiveness and computational efficiency. The parameters $\mu_b(\mathbf y)$ and $\sigma_b(\mathbf y)$ are predicted from the conditioning vector by a learned neural mapping. The conditional base distribution is parameterized as a diagonal Gaussian:
\begin{equation}\label{equ:base}
p_{\rm b}(\mathbf u\mid\mathbf y)
=\mathcal N\!\left(
\boldsymbol\mu_{\rm b}(\mathbf y),
\operatorname{diag}\!\left[\boldsymbol\sigma_{\rm b}^2(\mathbf y)\right]
\right).
\end{equation}
The same context is supplied to every autoregressive transform. Thus, the morphological descriptors control both the Gaussian base density and the transformation to the latent distribution.

The training of normalizing flows involves optimizing the parameters $\theta$ by maximizing the log-likelihood of the observed data. Following the change-of-variables formula, the conditional log-likelihood is computed as:
\begin{equation}\label{equ:log-likelihood}
\log p_\theta(\mathbf z\mid\mathbf y)
=
\log p_{\rm b}\!\left(f_\theta(\mathbf z;\mathbf y)\mid\mathbf y\right)
+
\log\left|
\det\frac{\partial f_\theta(\mathbf z;\mathbf y)}{\partial\mathbf z}
\right|.
\end{equation}
For image-derived latent vectors and their paired morphological contexts, the flow loss is expressed as the negative log-probability:
\begin{equation}\label{equ:flow_loss}
\mathcal L_{\rm flow}
=-\frac{1}{B}\sum_{i=1}^{B}
\log p_\theta(\mathbf z_i\mid\mathbf y_i).
\end{equation}

During the generation phase, the model draws $\mathbf u\sim p_{\rm b}(\mathbf u\mid\mathbf y)$, applies the inverse mapping $\mathbf z=f_\theta^{-1}(\mathbf u;\mathbf y)$, and generates the galaxy image $\widehat{\mathbf x}$ via the VAE decoder parameterizes $p_\psi(\mathbf{x}\mid\mathbf{z})$. Sampling the conditional base distribution allows different image realizations to be generated from the same prescribed morphological descriptors.

\subsection{Our Framework}\label{sec:framework}

\begin{figure*}
\centering
\includegraphics[width=\textwidth]{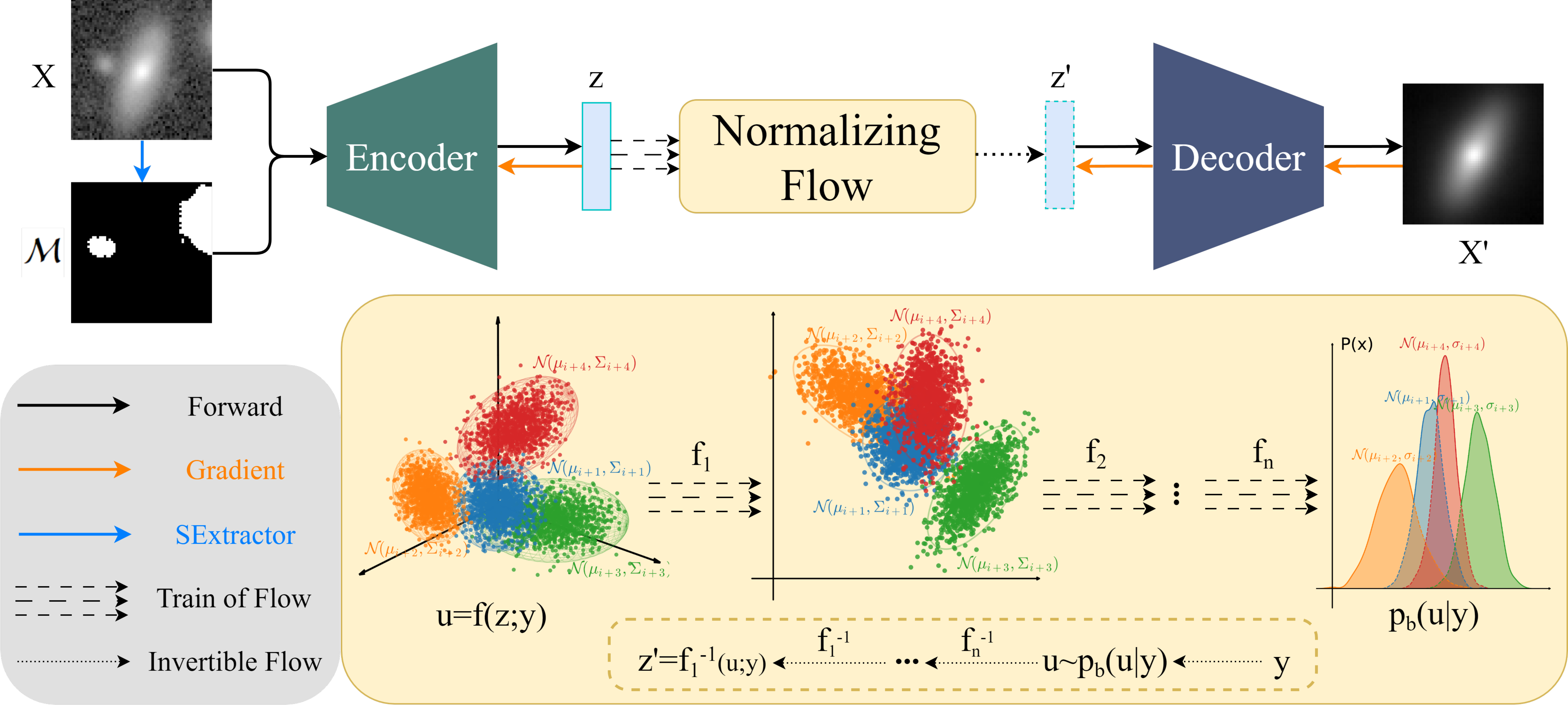}
\caption{Overview of the Flow-VAE framework. Galaxy images are encoded into VAE latent representations, while the non-parametric morphological descriptors condition the normalizing flow. Source masks are applied to the VAE reconstruction loss to reduce contamination from neighbouring sources.}
\label{fig:framework}
\end{figure*}

Figure~\ref{fig:framework} summarizes the proposed Flow-VAE framework for generating galaxy images conditioned on non-parametric morphological statistics. Each $r$-band galaxy image $\mathbf{x}$ is paired with its morphological descriptor vector $\mathbf{y}=(C,A,S,G,M_{20})$ and a source mask $\mathcal{M}$ derived from \texttt{SExtractor}. The VAE encoder maps the input image to a latent representation $\mathbf{z}$, while the conditional normalizing flow models the distribution $p_\theta(\mathbf{z}\mid\mathbf{y})$. In this formulation, the morphological descriptors act as conditioning variables that constrain the latent distribution rather than as direct outputs of the flow.

To reduce the influence of neighbouring sources, the reconstruction loss is evaluated over the unmasked pixels, which include the target galaxy and surrounding background. The masked reconstruction loss is defined as
\begin{equation}\label{equ:res_loss}
\mathcal{L}_{\text{rec}_\mathcal{M}} = \mathcal{M} \odot \| \mathbf{x}-\hat{\mathbf{x}} \|
\end{equation}
where $\hat{\mathbf{x}}$ is the reconstructed image and $\mathcal{M}$ is the binary source mask described in Section~\ref{sec:images}. This masking helps reduce the influence of contaminating sources when optimizing the reconstruction of the target galaxy.

The VAE and conditional flow are optimized to preserve both image fidelity and the statistical structure of the latent space. The overall objective is written as
\begin{equation}\label{equ:flow_vae_loss}
    \mathcal{L}_\text{Flow-VAE} = \mathcal{L}_{\text{rec}_\mathcal{M}} + \lambda_\text{flow} \cdot \mathcal{L}_\text{flow} + \lambda_\text{KL} \cdot \mathcal{L}_\text{KL}.
\end{equation}
where $\mathcal{L}_\text{KL}$ regularizes the VAE posterior toward its prior and $\mathcal{L}_\text{flow}$ maximizes the conditional likelihood of the image-derived latent vectors given the morphological descriptors. The coefficient $\lambda_\text{flow}$ controls the contribution of the conditional flow objective and is set to \num{0.1} in this work.

During generation, a target descriptor vector $\mathbf{y}$ is first specified and a base-space sample $\mathbf{u}$ is drawn from the corresponding conditional base distribution. The inverse flow maps this sample to a VAE latent vector, $\mathbf{z}=f_\theta^{-1}(\mathbf{u};\mathbf{y})$, which is then passed to the decoder to generate a galaxy image $\hat{\mathbf{x}}$. Because the sampling process is stochastic, multiple distinct galaxy images can be generated from the same prescribed morphological statistics, allowing the model to represent the intrinsic morphological diversity that is not captured by the five scalar descriptors alone. A practical implementation of generation from a complete five-dimensional descriptor vector is provided in Appendix~\ref{sec:code}.

We train the model using the Adam optimizer \citep{kingma2015iclradam} with a learning rate of 0.001, a batch size of 256, and 100 training epochs. The \num{240000} preprocessed $r$-band galaxy images are randomly divided into \num{216000} training images and \num{24000} test images. The model is implemented in PyTorch and trained on a single NVIDIA A100 GPU with 80 GB of memory.

\subsection{Gaussian Mixture Model for Parameter}\label{sec:gmm_prior}

Flow-VAE generates images from specified morphological descriptors, but its conditional accuracy may depend on how well the requested parameter combination is represented in the training data. We model their joint distribution using a Gaussian mixture model \citep[GMM,][]{Reynolds2009}, which represents the density of $\mathbf{y}=(C,A,S,G,M_{20})$ as
\begin{equation}\label{equ:gmm_joint}
p_{\rm GMM}(\mathbf y)
=\sum_{k=1}^{K}w_k
\mathcal N(\mathbf y\mid\boldsymbol\mu_k,\boldsymbol\Sigma_k),
\qquad
w_k\geq0,\quad \sum_{k=1}^{K}w_k=1,
\end{equation}
where $w_k$, $\boldsymbol\mu_k$, and $\boldsymbol\Sigma_k$ are the weight, mean, and covariance of component $k$. We adopt a GMM because it captures correlations and possible multimodality in the joint distribution while allowing analytic conditioning and density evaluation. These properties enable us to complete partially specified condition vectors using $p_{\rm GMM}(\mathbf{y}_{-j}\mid y_j)$ and to assess training-set support for requested parameter combinations.

This joint model supplies the missing conditions when only one descriptor is specified. Writing the fixed descriptor as $y_j$ and the remaining four as $\mathbf y_{-j}$, their conditional distribution is
\begin{equation}\label{equ:gmm_conditional}
p_{\rm GMM}(\mathbf y_{-j}\mid y_j)
=\frac{p_{\rm GMM}(\mathbf y)}{p_{\rm GMM}(y_j)}.
\end{equation}
For a GMM, conditioning gives another Gaussian mixture, with component weights, means, and covariances determined by the specified value. We sample $\mathbf y_{-j}$ from this distribution and combine it with $y_j$ to form a complete condition for Flow-VAE. Repeating the procedure keeps the selected descriptor fixed while allowing the others to vary according to the fitted correlations.

To assess the empirical support of these completed vectors and subsequently perturbed inputs, we evaluate their joint log-density:
\begin{equation}
s(\mathbf{y})
=\log p_{\rm GMM}(\mathbf{y})
=\log\left[
\sum_{k=1}^{K}w_k
\mathcal{N}(\mathbf{y}\mid\boldsymbol{\mu}_k,\boldsymbol{\Sigma}_k)
\right].
\end{equation}
The score accounts for the full parameter combination rather than its individual ranges. Section~\ref{sec:sample_ood} uses its empirical quantiles to examine generation errors in low-density regions. This density reflects the selected sample and its measurement procedure; low density indicates limited support under the fitted model, not physical impossibility.


\section{Results}\label{sec:results}

We apply the Flow-VAE to extract latent variables from the r-band DECaLS galaxy images and map them to physical parameters. Following \citet{10.1093/mnras/stad3181}, we adopt a 40-dimensional latent space, which optimally balances reconstruction fidelity, computational efficiency, and the training stability of the normalizing flow.

\subsection{Model Reconstruction Fidelity}\label{sec:reconstruction}

\begin{figure}
    \centering
    \includegraphics[width=\columnwidth]{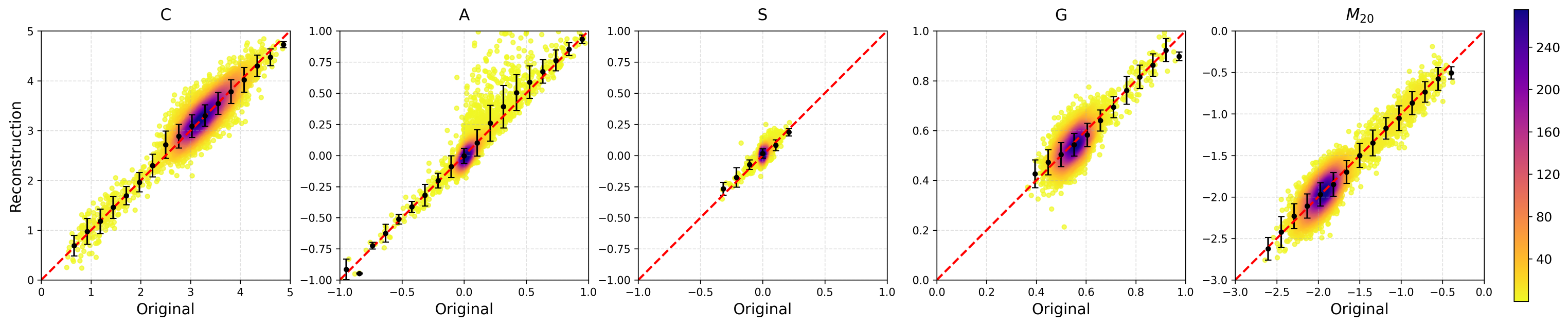}
    \caption{
    Comparison of non-parametric morphological statistics measured from original and VAE-reconstructed galaxy images.
    }
    \label{fig:recon}
\end{figure}

We evaluate reconstruction on the \num{24000}-image test set to assess how well the VAE preserves image structure and morphological statistics. Each image is encoded, passed through the VAE latent transformation, and decoded. The reconstructed images achieve a mean structural similarity index \citep[SSIM;][]{Wang2004} of 0.892, with values closer to 1 indicating greater similarity between the original and reconstructed images. To assess morphological fidelity, we also remeasure $C$, $A$, $S$, $G$, and $M_{20}$ using the same procedure applied to the original images. Figure~\ref{fig:recon} compares these measurements with their original values, with the identity line indicating exact agreement. For each bin of the original descriptor values, the figure also shows the median reconstructed value and the 16th--84th percentile interval, which corresponds to approximately $\pm1\sigma$ for a Gaussian distribution. These binned summaries reveal both systematic offsets from the identity relation and the scatter in the reconstructed measurements. No additional sky background or noise is added to the generated images before the morphological descriptors are measured.

We apply the same measurement procedure to original, reconstructed, and generated images, using identical SExtractor configurations and statmorph parameter settings. Because synthesized images may differ from the originals in both morphology and contaminating sources, we construct a new segmentation mask for each image rather than reuse the original mask. The detection containing the central pixel is treated as the target, while all other detections are masked before the morphological descriptors are measured.

To compare morphological fidelity across reconstruction and the subsequent generation experiments, we define the signed normalized residual for image $i$ and descriptor $j$ as
\begin{equation}\label{equ:error}
\delta_{ij}=\frac{\widehat y_{ij}-y_{ij}^{\rm ref}}{R_j},
\end{equation}
where $\widehat{y}_{ij}$ is the descriptor measured from the output image and $R_j$ is the fixed range width (see \ref{sec:morph_parameter} in detail) adopted for descriptor $j$ throughout the analysis. Normalization by $R_j$ enables comparison across descriptors without dividing by individual values that may be close to zero. Here, $y_{ij}^{\rm ref}$ is the descriptor value provided as input to the model, and it is measured from the original image, including the modified value in the perturbation experiment. SSIM and the descriptor recovery rates are calculated separately: SSIM is computed from the original and reconstructed image pairs, whereas the recovery rates quantify agreement between morphological descriptor values.



We define successful recovery as $|\delta_{ij}|\leq0.10$, corresponding to an absolute deviation within $10\%$ of the adopted descriptor range. Calculated over all valid reconstructed images, the recovery rates are $96.62\%$, $97.14\%$, $99.97\%$, $97.22\%$, and $96.96\%$ for $C$, $A$, $S$, $G$, and $M_{20}$, respectively. Thus, each descriptor is recovered within this tolerance in more than $96\%$ of valid reconstructions. Table~\ref{tab:success} also reports recovery rates at the stricter $5\%$ tolerance.

\subsection{Model Generation Capability}\label{sec:generation}

As illustrated in Figure~\ref{fig:framework}, Flow-VAE generates galaxy images from a complete condition vector $\mathbf{y}=(C,A,S,G,M_{20})$ using the procedure described in Appendix~\ref{sec:code}.
Figure~\ref{fig:conditional_examples} shows each observed galaxy alongside five images generated from its measured morphology statistics. The five descriptors do not uniquely determine the light distribution, as illustrated by the different major-axis orientations in the bottom row. We therefore assess conditional fidelity by comparing the requested descriptor values with those measured from the generated images.

\begin{figure}
    \centering
    \includegraphics[width=0.6\columnwidth]{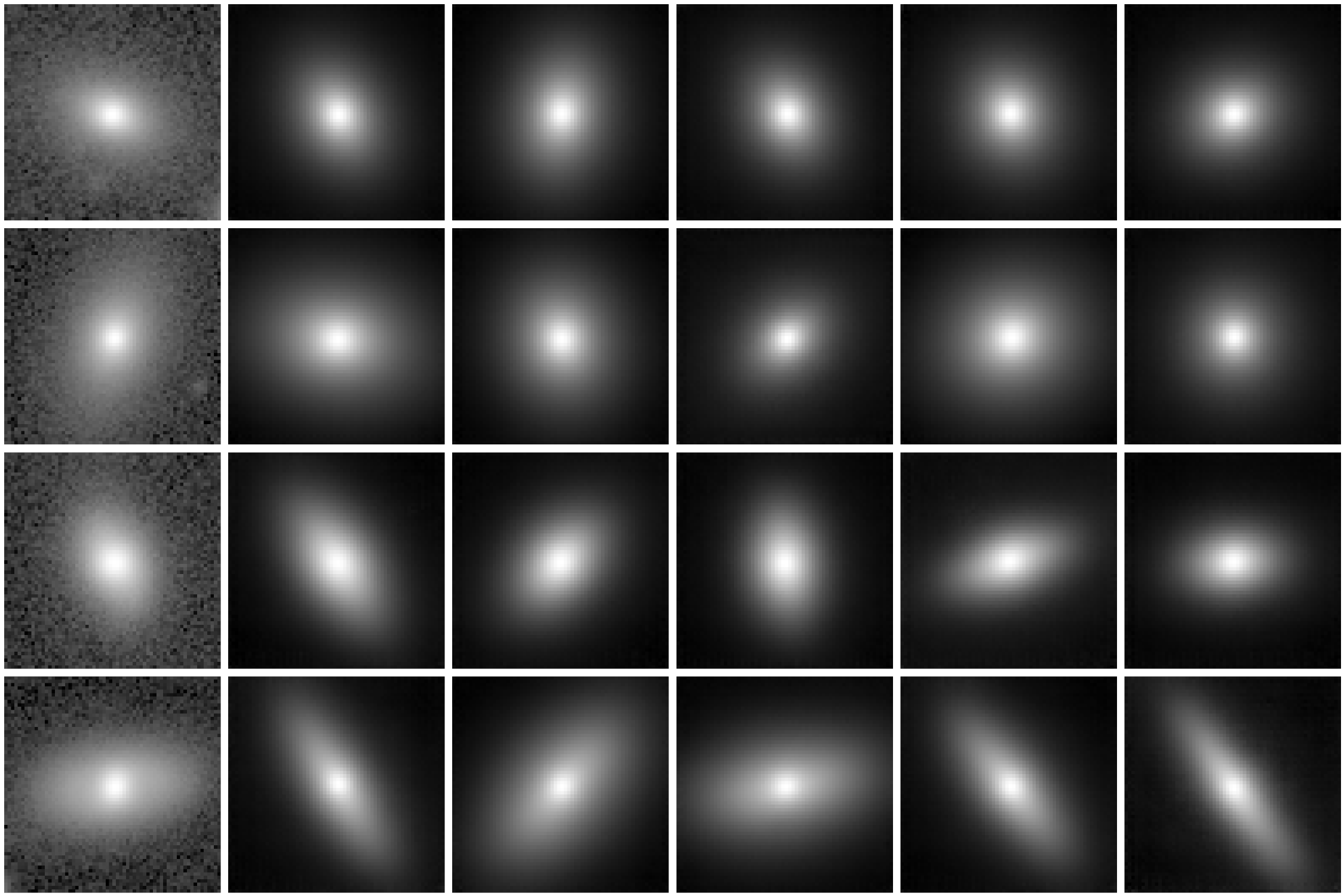}
    \caption{Observed test galaxies and Flow-VAE images generated using the five morphological descriptors ($C$, $A$, $S$, $G$, and $M_{20}$).}
    \label{fig:conditional_examples}
\end{figure}


We evaluate conditional generation fidelity from two complementary perspectives. First, Section~\ref{sec:single_generation} quantifies descriptor-specific control by fixing one of the five statistics and sampling the remaining four from their GMM conditional distribution. This experiment measures how accurately the model recovers each prescribed descriptor while retaining statistically supported combinations of the other morphological properties. Section~\ref{sec:sample_ood} evaluates complete condition vectors and compares their generation errors across different GMM likelihoods. The GMM likelihood measures training-distribution support, allowing us to test whether joint conditional fidelity decreases for less probable parameter combinations.

\subsubsection{Single-parameter Conditional Generation}\label{sec:single_generation}

We first perform a single-parameter conditional-generation experiment. For a target descriptor $y_j$ spanning its test-set distribution, the specified value is held fixed while the remaining four descriptors, denoted by $\mathbf{y}_{-j}$, are sampled from the GMM conditional distribution in Equation~\ref{equ:gmm_conditional}. The completed five-dimensional vectors are then supplied to Flow-VAE to generate an ensemble of galaxy images for each target value. Single-parameter generation is a special case of the more general partial-parameter procedure, which supports any specified subset of the five descriptors and is described in Appendix~\ref{sec:code1}.

\begin{figure}
    \centering
	\includegraphics[width=\columnwidth]{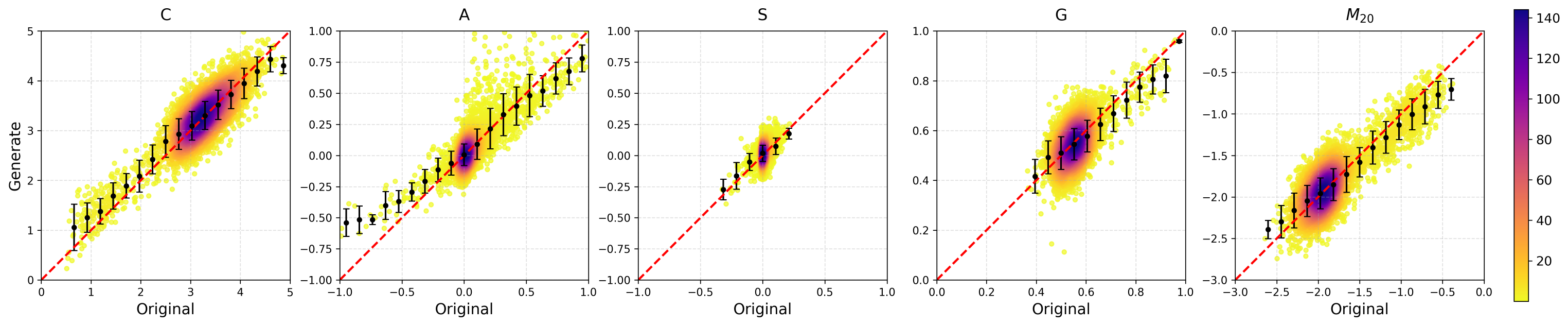}
    \caption{
    Comparison of Non-parametric morphological Statistics between observed and single-parameter generated galaxies.
    }
    \label{fig:single_generation}
\end{figure}



\begin{table}[h]
	\centering
	\caption{Morphological descriptor recovery rates for reconstruction, single-parameter generation, and a constant-median baseline. $R_j$ denotes the adopted descriptor range width, and $\delta_{ij}$ is defined in Equation~\ref{equ:error}. For each descriptor, the baseline always predicts its median in the test set, independently of the input value. }
	\label{tab:success}
    \setlength{\tabcolsep}{1.2mm}{
	\begin{tabular}{lccccccc}
        \toprule
        \multirow{2}{*}{Descriptor} & \multirow{2}{*}{$R_j$} & \multicolumn{2}{c}{Reconstruction} & \multicolumn{2}{c}{Generation} & \multicolumn{2}{c}{Null model (median)} \\
\cmidrule(lr){3-4}\cmidrule(lr){5-6}\cmidrule(lr){7-8}
 &  & $|\delta|\leq0.05$ & $|\delta|\leq0.10$ & $|\delta|\leq0.05$ & $|\delta|\leq0.10$ & $|\delta|\leq0.05$ & $|\delta|\leq0.10$ \\
\midrule
$C$      & 4.00 & 75.2\% & 96.62\% & 61.6\% & 90.95\% & 39.37\% & 69.09\% \\
$A$      & 1.70 & 91.3\% & 97.14\% & 72.1\% & 94.92\% & 65.04\% & 88.74\% \\
$S$      & 0.45 & 98.5\% & 99.97\% & 88.7\% & 99.83\% & 77.39\% & 87.49\% \\
$G$      & 0.55 & 72.7\% & 97.22\% & 56.3\% & 87.81\% & 47.11\% & 78.58\% \\
$M_{20}$ & 2.00 & 74.2\% & 96.96\% & 56.2\% & 88.79\% & 48.43\% & 75.74\% \\
        \bottomrule
	\end{tabular}}
\end{table}

To assess single-parameter generation fidelity, we remeasure the specified descriptor from each synthesized image and compare it with its requested value (Figure~\ref{fig:single_generation}). Table~\ref{tab:success} reports recovery rates at both the $5\%$ and $10\%$ tolerances, together with the constant-median baseline. Reconstruction and generation exceed this baseline for all five descriptors at both tolerances. For $S$, however, the baseline alone achieves recovery rates of $77.39\%$ and $87.49\%$ at the $5\%$ and $10\%$ tolerances, respectively, compared with $88.7\%$ and $99.83\%$ for conditional generation. These baseline rates show that the concentration of descriptor values can produce substantial recovery without input-dependent predictions. The high recovery rate of $S$ alone should therefore not be interpreted as evidence of stronger conditional control than for the other descriptors.

The binned input--output relations show regression toward the center of the training distribution: low requested values tend to be overestimated, whereas high values tend to be underestimated, particularly for $C$, $A$, and $G$.
This compression qualitatively resembles the attenuation of extreme predictions discussed by \citet{2025OJAp....8E..95T}; however, the present experiment does not establish whether measurement-error-induced attenuation is responsible for the trend.





\subsubsection{Joint-parameter Conditional Generation}\label{sec:sample_ood}

Here, we focus on complete condition vectors, $\mathbf{y}=(C,A,S,G,M_{20})$, and examine whether their generation fidelity depends on their representation in the joint training distribution. Because the five descriptors are correlated, a condition vector can lie within the adopted range of every individual descriptor while remaining rare as a five-dimensional combination. We quantify this training-set support using the GMM log-density score ($s=\log p(\mathbf{y})$).


Because the distribution of $s$ is not generally Gaussian, we convert its empirical percentile rank into a Gaussian-equivalent scale:
\begin{equation}
n_\sigma(s)=-\Phi^{-1}\!\left[\widehat{F}_s(s)\right], 
\end{equation}
where $\widehat{F}_s(s)$ is the empirical cumulative distribution function of the training-set scores and $\Phi^{-1}$ is the inverse standard normal cumulative distribution function. Larger $n_\sigma$ values therefore indicate condition vectors with lower GMM density and weaker representation in the training sample. In particular, $n_\sigma>3$ corresponds to the lowest approximately $0.135\%$ of the training-score distribution. We use this threshold as a descriptive reference for the low-density tail.

\begin{figure}
    \includegraphics[width=\columnwidth]{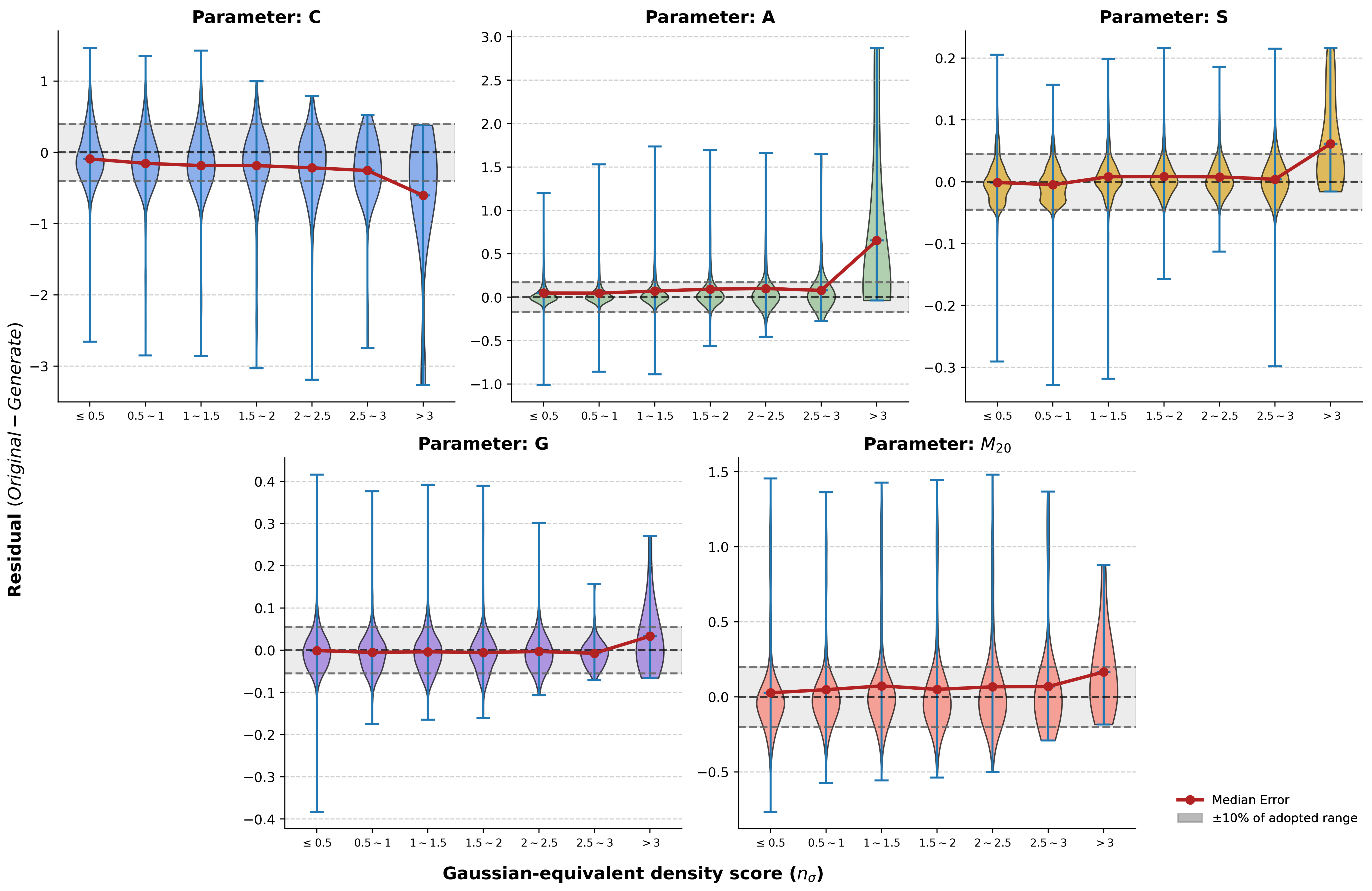}
    \caption{
    Residual distributions of the generated morphological statistics as a function of GMM input-density score. The violin shapes represent error densities within discrete $0.5\sigma$ bins, while the connected line traces the median residual in each bin. 
    }
    \label{fig:gmm_residual}
\end{figure}

To probe the model under progressively less supported conditions, we construct a bounded perturbation sample from the test set. For each condition vector, we randomly select one descriptor and perturb it by up to $10\%$ of its adopted range, while keeping the other four descriptors fixed. The perturbed value is clipped to the corresponding training-set range, ensuring that every descriptor remains marginally valid. Nevertheless, changing one descriptor independently can weaken its consistency with the other four and move the complete condition vector into a lower-density region of the joint distribution. 

Figure~\ref{fig:gmm_residual} shows that the residual distributions generally broaden as the condition vectors move toward larger $n_\sigma$. Several descriptors also develop systematic offsets in the low-density regime, indicating that the generated measurements are increasingly drawn toward combinations better represented in the training sample. Conditional fidelity therefore depends on the joint support of the complete input vector, rather than only on whether each descriptor lies within its individual training range.

These results show that the model can accept bounded combinations of all five descriptors, but its conditional fidelity is highest for combinations well represented in the training data and decreases toward lower-density regions. The GMM density score therefore provides a useful measure of training-set support for a requested condition, but it does not represent the probability of successful generation or determine whether the corresponding morphology is physically feasible. Even in the low-density regime $(n_\sigma>3)$, 22.5\% of valid generated images recover all five requested descriptors within 10\% of their adopted ranges. This indicates some capacity to generalize to sparsely represented combinations within the tested perturbation setting, despite the decline in conditional fidelity.


\section{Discussion}\label{sec:discussion}

Flow-VAE extends previous galaxy-generation methods conditioned mainly on magnitude, size, and S{\'e}rsic index by conditioning image generation on non-parametric morphological descriptors \citep{Lanusse2021, Euclid}. Under the same recovery criterion, single-parameter generation yields descriptor recovery rates $0.14$--$9.41$ percentage points below those of reconstruction. This difference shows that preserving morphology during reconstruction does not guarantee equally accurate conditional generation. Smoothness has the highest generation recovery rate and the smallest reduction from reconstruction, while $G$ and $M_{20}$ have the lowest recovery rates.

The lower fidelity of $G$ and $M_{20}$ may reflect their sensitivity to pixel-level flux variations and, for $M_{20}$, the spatial distribution of bright regions, which may not be fully captured by the adopted 40-dimensional latent space. Related work shows that high reconstruction fidelity does not necessarily imply readily accessible physical information in the learned representation \citep{Sanjaripour2026}. Increasing the latent dimensionality could improve morphological fidelity but may also complicate conditional density modeling with normalizing flows. A systematic investigation of this trade-off is beyond the scope of this work.

\begin{figure}
    \includegraphics[width=\columnwidth]{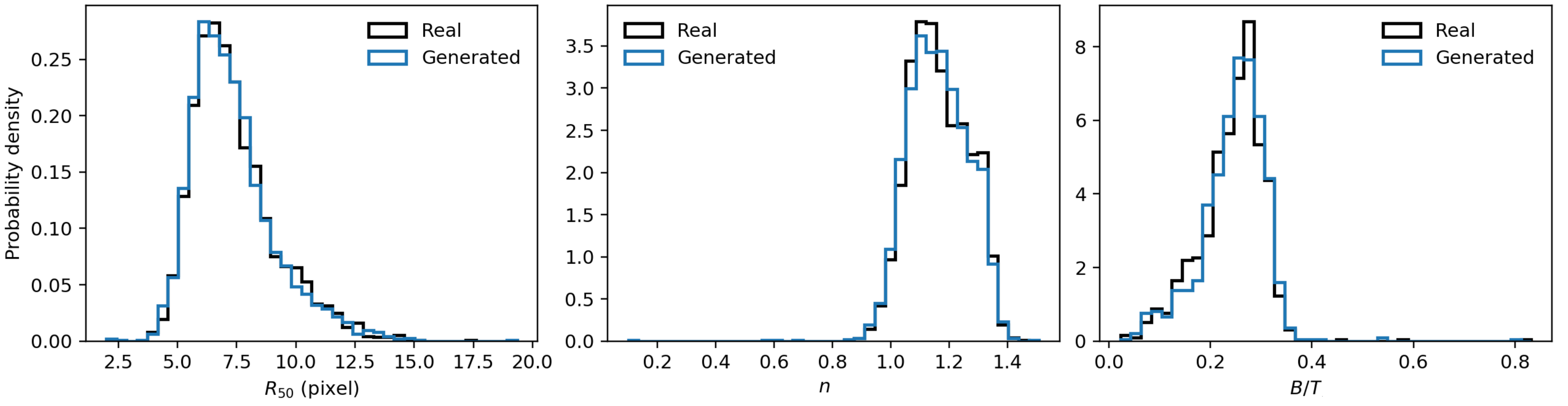}
    \caption{
    Distributions of the circular half-light radius $R_{50}$ (left), effective single-S\'ersic index $n$ (middle), and model-based bulge-to-total light ratio $B/T$ (right) for real (black) and generated (blue) images.
    }
    \label{fig:independent_properties}
\end{figure}

To complement the visual comparison in Section~\ref{sec:generation}, we examined how closely the conditionally generated sample reproduces the observed distributions of three structural properties beyond the five conditioning descriptors (Figure~\ref{fig:independent_properties}). We applied the same measurement procedure to both samples without PSF correction. We defined $R_{50}$ as the circular radius enclosing half the flux measured within $1.5$ times the circular Petrosian radius. A single S\'ersic fit provided the effective index $n$, while a fit with a S\'ersic bulge and an exponential disc provided $B/T$, the fraction of total model light assigned to the bulge. The marginal distributions of $R_{50}$, $n$, and $B/T$ show close agreement, with Kolmogorov--Smirnov distances of $0.023$, $0.022$, and $0.038$, respectively. Both samples peak in apparent galaxy size at $R_{50}\simeq 6$--$7$ pixels, with fitted S\'ersic indices just above unity, indicating similar characteristic sizes and predominantly near-exponential light profiles at the observed resolution. The largest of the three distances occurs for $B/T$, whose distribution shows a slightly sharper peak near $0.27$ in the observed sample.

Flow-VAE is trained on observed $r$-band images from the DESI Legacy Imaging Surveys, with the five conditioning descriptors measured from those same images. The descriptor recovery rates therefore quantify agreement with morphology at the resolution of the observations. Effective image resolution also affects the interpretation of the additional structural measurements, particularly how well the central and outer light components can be distinguished when estimating $B/T$. The fixed $64\times64$ field of view also limits the applicability to extended nearby galaxies, for which the cutout may truncate the outer light distribution or the Petrosian measurement aperture and leave insufficient source-free area for the background corrections to $A$ and $S$. For small, weakly resolved galaxies, the measured $A$, $S$, and $M_{20}$ remain sensitive to PSF smoothing, whose physical scale varies with redshift.

The masked MAE reconstruction loss measures absolute pixel differences over unmasked pixels, without the inverse-variance weighting used in a $\chi^2$ objective \citep{universe12090285}. It does not explicitly account for spatial variations in noise uncertainty, so the reconstruction scores alone do not establish how accurately the model recovers the underlying noise-free signal. The training sample is dominated by low-redshift ($z<0.1$) galaxies; applications to other surveys, bands, or redshift ranges therefore need validation with representative data, particularly where spatial resolution and noise properties differ from those of the training sample. Nevertheless, the framework is expected to be transferable to these other surveys, subject to appropriate retraining and validation on representative data.

\section{Conclusions}\label{sec:conclusions}

We present Flow-VAE, a framework for generating galaxy images conditioned on five non-parametric morphological statistics: $C$, $A$, $S$, $G$, and $M_{20}$. The model combines a variational autoencoder with a conditional normalizing flow, while a GMM represents the joint distribution of the conditioning variables. This design allows any subset of the five descriptors to be prescribed, with the remaining descriptors sampled conditionally to avoid unsupported independent combinations.

The main results are as follows:
\begin{itemize}
\item \textbf{Model Reconstruction Fidelity:} The VAE achieves a mean SSIM of $0.892$ on the \num{24000}-image test set. Using $|\delta_{ij}|\leq0.10$ as the recovery criterion, the reconstruction success rates are $96.62\%$ for $C$ , $97.14\%$ for $A$, $99.97\%$ for $S$, $97.22\%$ for $G$, and $96.96\%$ for $M_{20}$. 

\item \textbf{Single-parameter morphological control:} Conditional GMM sampling enables generation with one prescribed descriptor while accounting for the empirical relationships among the remaining four. The generated measurements broadly follow the requested values, although systematic departures remain toward the ends of the sampled ranges.

\item \textbf{Training-distribution support:} The joint GMM density provides an empirical measure of generation reliability across the five-dimensional morphological parameter space. The Gaussian-equivalent $3\sigma$ boundary marks the input regions where conditional performance has been evaluated.
\end{itemize}

The principal limitations are systematic regression toward common descriptor values and reduced conditional accuracy for sparsely represented parameter combinations. These effects should be considered when generating samples with rare or extreme morphological statistics. Within well-supported regions of the training distribution, Flow-VAE provides a practical method for generating galaxy samples with controlled morphological descriptors without imposing analytical surface-brightness profiles. However, agreement in these descriptors does not by itself establish visual realism or the recovery of specific morphological structures. Subject to validation for the intended application, the generated samples could support survey-pipeline tests, selection-effect studies, and assessments of morphology-dependent biases in surveys such as Euclid and CSST.

The current conditioning scheme does not directly specify morphological classes or individual structures such as bars, spiral arms, rings, or tidal features. Agreement in the five descriptors alone does not establish visual realism, and the conditional distributions of size, axis ratio, and S{\'e}rsic index remain unvalidated.

\funding{This work was supported by the National Key R\&D Program of China (Nos.2025YFF0510603), the National Natural Science Foundation of China (Nos.12141302), and the China Manned Space Project (CMS-CSST-2025-A07).}

\dataavailability{This work uses the galaxy image (DECaLS) data underlying this article are available at \url{https://www.legacysurvey.org/}. Our code is publicly available: \url{https://github.com/xuquanfeng/Galaxy_Flow_VAE}.} 

\acknowledgments{The DESI Legacy Imaging Surveys consist of three complementary projects: the Dark Energy Camera Legacy Survey (DECaLS), Beijing–Arizona Sky Survey (BASS), and Mayall z-band Legacy Survey (MzLS). DECaLS, BASS, and MzLS include data obtained, respectively, with the Blanco telescope at the Cerro Tololo Inter-American Observatory (CTIO), operated by NSF’s National Optical-Infrared Astronomy Research Laboratory (NOIRLab); the Bok telescope at Steward Observatory; and the Mayall telescope at Kitt Peak National Observatory (KPNO), operated by NOIRLab. Pipeline processing was supported by NOIRLab and Lawrence Berkeley National Laboratory (LBNL). This project used data obtained with the Dark Energy Camera (DECam), constructed by the Dark Energy Survey (DES) collaboration; funding for DES was provided by the U.S. Department of
Energy (DOE), the National Science Foundation (NSF), and numerous international agencies (full list at \url{https://www.darkenergysurvey.org}). BASS is a key project of the Telescope Access Program (TAP), funded by the National Astronomical Observatories, Chinese Academy of Sciences (NAOC), and the Chinese Academy of Sciences. The Legacy Surveys team thanks the Tohono O’odham Nation for access to Iolkam Du’ag (Kitt Peak). NOIRLab is operated by the Association of Universities for Research in Astronomy (AURA) under a cooperative agreement with NSF; LBNL is managed by the Regents of the University of California under contract to the U.S. Department of Energy. The Legacy Surveys imaging of the DESI footprint is supported by DOE Contract No. DE-AC02-05CH1123 and
NSF Contract No. AST-0950945.
}

\conflictsofinterest{The authors declare no conflicts of interest.} 

\appendixtitles{yes} 
\appendixstart
\appendix

\section[\appendixname~\thesection]{Generation from Complete Morphological Conditions}\label{sec:code}

This appendix describes conditional galaxy generation when all five non-parametric morphological statistics are specified. The user supplies a complete condition vector $\mathbf{y}=(C,A,S,G,M_{20})$. Before generation, the pre-fitted GMM evaluates the joint density of the requested vector, providing an empirical indication of how well the condition is represented in the training sample. This density score characterizes statistical support and should not be interpreted as a probability of successful generation or as a criterion of physical feasibility.

Given the complete condition vector, a base-space latent variable $\mathbf{u}$ is sampled from the conditional flow distribution and transformed through the inverse flow as
\begin{equation}
\mathbf{z}=f_{\theta}^{-1}(\mathbf{u};\mathbf{y}).
\end{equation}
The resulting latent vector $\mathbf{z}$ is passed to the VAE decoder to generate a galaxy image. Repeated sampling at the same $\mathbf{y}$ produces different realizations associated with the same requested morphological conditions. Because the five statistics do not uniquely specify an image, these realizations are not expected to reproduce a particular observed galaxy at the pixel level. Conditional fidelity should instead be evaluated by remeasuring the five statistics from the generated images and comparing them with their requested values.

\begin{lstlisting}[language=Python]
import joblib
import numpy as np
import GMMPercentileScorer
import torch
import Flow_VAE
state = joblib.load('gmm_state.pkl')
scorer = GMMPercentileScorer(n_components=500)
scorer.load_state(state)
y = np.array([3.05, 0.02, 0.01, 0.52, -1.65])
res = scorer.score_point(y)
print("percentile %:", res['percentile_pct'])
# percentile %: 81.2210939945057
model = Flow_VAE()
model.load_state_dict(torch.load('Flow_VAE.pth'))
u = model.sample_base(condition=y)
z = model.flow_inverse(u,condition=y)
imgs = model.decode(z)
\end{lstlisting}

\section[\appendixname~\thesection]{Generation from Partially Specified Morphological Conditions}\label{sec:code1}

This appendix describes conditional generation when only a subset of the five morphological statistics is specified. Let $\mathbf{y}_{\mathcal{K}}$ denote the specified descriptors and $\mathbf{y}_{\mathcal{U}}$ the unspecified descriptors. The pre-fitted GMM is used to derive the conditional distribution
\begin{equation}
p_{\rm GMM}(\mathbf{y}_{\mathcal{U}}\mid\mathbf{y}_{\mathcal{K}}),
\end{equation}
from which the missing values are sampled. Combining the specified and sampled values produces a complete five-dimensional condition vector that preserves the correlations represented in the training population.

The completed vector is subsequently passed to Flow-VAE using the complete-condition generation procedure described in Appendix~\ref{sec:code}. Specifying one descriptor and sampling the remaining four corresponds to the single-parameter experiment in Section~\ref{sec:single_generation}; the same procedure also supports any other subset of the five descriptors.

\begin{lstlisting}[language=Python]
import cal_gmm
import sample_gmm
# Input
know_idx = [0, 2] # Assuming known x0, x2
know = [3.22, -0.06]
num_missing = 3

gmm = state['gmm']
miss_idx = [i for i in range(5) if i not in known_idx]
w, means, covs = cal_gmm(gmm, know, know_idx, miss_idx)
samples = sample_gmm(w, means, covs, n_samples=1)
final = np.zeros(5)
final[know_idx] = x_known
final[missing_idx] = samples
print(', '.join([f'{final[i]}' for i in range(5)]))
# 3.22, -0.28, -0.06, 0.53, -1.72
\end{lstlisting}

\begin{adjustwidth}{-\extralength}{0cm}

\reftitle{References}


\externalbibliography{yes}
\bibliography{references}

@article{Robertson2019GalaxyFA,
  title={Galaxy formation and evolution science in the era of the Large Synoptic Survey Telescope},
  author={Brant E. Robertson and Manda Banerji and Sarah Brough and Roger L. Davies and Henry C. Ferguson and Ryan Hausen and Sugata Kaviraj and Jeffrey A. Newman and Samuel J. Schmidt and J. Anthony Tyson and Risa Wechsler},
  journal={Nature Reviews Physics},
  year={2019},
  volume={1},
  pages={450 - 462},
  url={https://api.semanticscholar.org/CorpusID:257115136}
}

@article{10.1093/mnras/stad3181,
    author = {Xu, Quanfeng and Shen, Shiyin and de Souza, Rafael S and Chen, Mi and Ye, Renhao and She, Yumei and Chen, Zhu and Ishida, Emille E O and Krone-Martins, Alberto and Durgesh, Rupesh},
    title = "{From images to features: unbiased morphology classification via variational auto-encoders and domain adaptation}",
    journal = {Monthly Notices of the Royal Astronomical Society},
    volume = {526},
    number = {4},
    pages = {6391-6400},
    year = {2023},
    month = {10},
    issn = {0035-8711},
    doi = {10.1093/mnras/stad3181},
    url = {https://doi.org/10.1093/mnras/stad3181},
    eprint = {https://academic.oup.com/mnras/article-pdf/526/4/6391/52662489/stad3181.pdf},
}

@article{10.1126/science.1060855,
author = {Roberto G. Abraham  and Sidney van den Bergh },
title = {The Morphological Evolution of Galaxies},
journal = {Science},
volume = {293},
number = {5533},
pages = {1273-1278},
year = {2001},
doi = {10.1126/science.1060855},
URL = {https://www.science.org/doi/abs/10.1126/science.1060855},
eprint = {https://www.science.org/doi/pdf/10.1126/science.1060855}}

@inproceedings{DinhKB14,
  author={Laurent Dinh and David Krueger and Yoshua Bengio},
  title={NICE: Non-linear Independent Components Estimation},
  year={2015},
  cdate={1420070400000},
  url={http://arxiv.org/abs/1410.8516},
  booktitle={ICLR (Workshop)}
}

@inproceedings{DurkanB0P19,
  author={Conor Durkan and Artur Bekasov and Iain Murray and George Papamakarios},
  title={Neural Spline Flows},
  year={2019},
  cdate={1546300800000},
  pages={7509-7520},
  url={http://papers.nips.cc/paper/8969-neural-spline-flows},
  booktitle={NeurIPS}
}

@article{Lanusse2021,
    author = {Lanusse, François and Mandelbaum, Rachel and Ravanbakhsh, Siamak and Li, Chun-Liang and Freeman, Peter and Póczos, Barnabás},
    title = "{Deep generative models for galaxy image simulations}",
    journal = {Monthly Notices of the Royal Astronomical Society},
    volume = {504},
    number = {4},
    pages = {5543-5555},
    year = {2021},
    month = {05},
    issn = {0035-8711},
    doi = {10.1093/mnras/stab1214},
    url = {https://doi.org/10.1093/mnras/stab1214},
    eprint = {https://academic.oup.com/mnras/article-pdf/504/4/5543/38036124/stab1214.pdf},
}

@article{2025OJAp....8E..95T,
       author = {{Ting}, Yuan-Sen},
        title = "{Why Machine Learning Models Systematically Underestimate Extreme Values}",
      journal = {The Open Journal of Astrophysics},
         year = 2025,
        month = 7,
       volume = {8},
          eid = {95},
        pages = {95},
          doi = {10.33232/001c.142224},
archivePrefix = {arXiv},
       eprint = {2412.05806},
 primaryClass = {astro-ph.IM}
}

@Article{universe12090285,
AUTHOR = {Ye, Renhao and Shen, Shiyin and Xu, Quanfeng},
TITLE = {From Observation to Ground Truth: Reconstruction Losses of Galaxy Imaging Under Heteroscedastic Noise},
JOURNAL = {Universe},
VOLUME = {12},
YEAR = {2026},
NUMBER = {9},
ARTICLE-NUMBER = {285},
URL = {https://www.mdpi.com/2218-1997/12/9/285},
ISSN = {2218-1997},
DOI = {10.3390/universe12090285}
}

@article{10.1093/mnras/stae1577,
    author = {He, Qiuhan and Nightingale, James W and Amvrosiadis, Aris and Robertson, Andrew and Cole, Shaun and Frenk, Carlos S and Massey, Richard and Li, Ran and Cao, Xiaoyue and Lange, Samuel C and França, João Paulo C},
    title = {Unveiling lens light complexity with a novel multi-Gaussian expansion approach for strong gravitational lensing},
    journal = {Monthly Notices of the Royal Astronomical Society},
    volume = {532},
    number = {2},
    pages = {2441-2462},
    year = {2024},
    month = {08},
    issn = {0035-8711},
    doi = {10.1093/mnras/stae1577},
    url = {https://doi.org/10.1093/mnras/stae1577},
    eprint = {https://academic.oup.com/mnras/article-pdf/532/2/2441/58559110/stae1577.pdf},
}

@article{10.1093/mnras/stw3039,
    author = {Robotham, A. S. G. and Taranu, D. S. and Tobar, R. and Moffett, A. and Driver, S. P.},
    title = {ProFit: Bayesian profile fitting of galaxy images},
    journal = {Monthly Notices of the Royal Astronomical Society},
    volume = {466},
    number = {2},
    pages = {1513-1541},
    year = {2017},
    month = {04},
    issn = {0035-8711},
    doi = {10.1093/mnras/stw3039},
    url = {https://doi.org/10.1093/mnras/stw3039},
    eprint = {https://academic.oup.com/mnras/article-pdf/466/2/1513/10867054/stw3039.pdf},
}

@article{10.1111/j.1365-2966.2010.17690.x,
    author = {Andrae, René and Jahnke, Knud and Melchior, Peter},
    title = {Parametrizing arbitrary galaxy morphologies: potentials and pitfalls},
    journal = {Monthly Notices of the Royal Astronomical Society},
    volume = {411},
    number = {1},
    pages = {385-401},
    year = {2011},
    month = {01},
    issn = {0035-8711},
    doi = {10.1111/j.1365-2966.2010.17690.x},
    url = {https://doi.org/10.1111/j.1365-2966.2010.17690.x},
    eprint = {https://academic.oup.com/mnras/article-pdf/411/1/385/3517524/mnras0411-0385.pdf},
}

@misc{Papamakarios2017,
      title={Masked Autoregressive Flow for Density Estimation}, 
      author={George Papamakarios and Theo Pavlakou and Iain Murray},
      year={2018},
      eprint={1705.07057},
      archivePrefix={arXiv},
      primaryClass={stat.ML},
      url={https://arxiv.org/abs/1705.07057}, 
}

@article{kingma2013auto,
  title={Auto-encoding variational bayes},
  author={Kingma, Diederik P and Welling, Max},
  journal={arXiv preprint arXiv:1312.6114},
  year={2013}
}

@article{Kullback1951,
author = "Kullback, S. and Leibler, R. A.",
doi = "10.1214/aoms/1177729694",
fjournal = "The Annals of Mathematical Statistics",
journal = "Ann. Math. Statist.",
month = "03",
number = "1",
pages = "79--86",
publisher = "The Institute of Mathematical Statistics",
title = "On Information and Sufficiency",
url = "http://dx.doi.org/10.1214/aoms/1177729694",
volume = "22",
year = "1951"
}

@article{Dey_2019,
	doi = {10.3847/1538-3881/ab089d},
	url = {https://doi.org/10.3847/1538-3881/ab089d},
	year = 2019,
	month = {apr},
	publisher = {American Astronomical Society},
	volume = {157},
	number = {5},
	pages = {168},
	author = {Arjun Dey and David J. Schlegel and Dustin Lang and Robert Blum and Kaylan Burleigh},
	title = {Overview of the {DESI} Legacy Imaging Surveys},
	journal = {The Astronomical Journal}
}

@misc{Sanjaripour2026,
      title={The Galaxy's Guide to the Tokenizer: A Benchmark for Scientific Foundation Models}, 
      author={Sogol Sanjaripour and Michael J. Smith and Manuel Pérez-Carrasco and Juan Rafael Martínez-Galarza and Bahram Mobasher and Gabriela Canalizo},
      year={2026},
      eprint={2606.25610},
      archivePrefix={arXiv},
      primaryClass={astro-ph.IM},
      url={https://arxiv.org/abs/2606.25610}, 
}

@article{refId0,
	author = {{Bertin, E.} and {Arnouts, S.}},
	title = {SExtractor: Software for source extraction},
	DOI= "10.1051/aas:1996164",
	url= "https://doi.org/10.1051/aas:1996164",
	journal = {Astron. Astrophys. Suppl. Ser.},
	year = 1996,
	volume = 117,
	number = 2,
	pages = "393-404",
}

@article{10.1093/mnras/sty3345,
    author = {Rodriguez-Gomez, Vicente and Snyder, Gregory F and Lotz, Jennifer M and Nelson, Dylan and Pillepich, Annalisa and Springel, Volker and Genel, Shy and Weinberger, Rainer and Tacchella, Sandro and Pakmor, Rüdiger and Torrey, Paul and Marinacci, Federico and Vogelsberger, Mark and Hernquist, Lars and Thilker, David A},
    title = "{The optical morphologies of galaxies in the IllustrisTNG simulation: a comparison to Pan-STARRS observations}",
    journal = {Monthly Notices of the Royal Astronomical Society},
    volume = {483},
    number = {3},
    pages = {4140-4159},
    year = {2018},
    month = {12},
    issn = {0035-8711},
    doi = {10.1093/mnras/sty3345},
    url = {https://doi.org/10.1093/mnras/sty3345},
    eprint = {https://academic.oup.com/mnras/article-pdf/483/3/4140/27330457/sty3345.pdf},
}

@article{10.1093/mnras/stt1016,
    author = {Freeman, P. E. and Izbicki, R. and Lee, A. B. and Newman, J. A. and Conselice, C. J. and Koekemoer, A. M. and Lotz, J. M. and Mozena, M.},
    title = "{New image statistics for detecting disturbed galaxy morphologies at high redshift}",
    journal = {Monthly Notices of the Royal Astronomical Society},
    volume = {434},
    number = {1},
    pages = {282-295},
    year = {2013},
    month = {06},
    issn = {0035-8711},
    doi = {10.1093/mnras/stt1016},
    url = {https://doi.org/10.1093/mnras/stt1016},
    eprint = {https://academic.oup.com/mnras/article-pdf/434/1/282/18499355/stt1016.pdf},
}

@article{Conselice_2003,
doi = {10.1086/375001},
url = {https://dx.doi.org/10.1086/375001},
year = {2003},
month = {jul},
publisher = {},
volume = {147},
number = {1},
pages = {1},
author = {Christopher J. Conselice},
title = {The Relationship between Stellar Light Distributions of Galaxies and
Their Formation Histories},
journal = {The Astrophysical Journal Supplement Series}
}

@article{Lotz_2004,
doi = {10.1086/421849},
url = {https://dx.doi.org/10.1086/421849},
year = {2004},
month = {jul},
publisher = {},
volume = {128},
number = {1},
pages = {163},
author = {Jennifer M. Lotz and Joel Primack and Piero Madau},
title = {A New Nonparametric Approach to Galaxy Morphological Classification},
journal = {The Astronomical Journal}
}

@Inbook{Reynolds2009,
  title={Gaussian mixture models.},
  author={Reynolds, Douglas},
  journal={Encyclopedia of biometrics},
  volume={741},
  number={659-663},
  pages={3},
  isbn={978-0-387-73003-5},
  year={2009},
  publisher={Springer US}
}

@inproceedings{kingma2015iclradam,
  title     = {{Adam: A Method for Stochastic Optimization}},
  author    = {Kingma, Diederik P. and Ba, Jimmy},
  booktitle = {International Conference on Learning Representations},
  year      = {2015},
  url       = {https://mlanthology.org/iclr/2015/kingma2015iclr-adam/}
}

@article{Euclid,
	author = {{Euclid Collaboration} and {Bretonnière, H.} and {Huertas-Company, M.} and {Boucaud, A.} and {Lanusse, F.} and {Jullo, E.} and {Merlin, E.} and {Tuccillo, D.} and {Castellano, M.} and {Brinchmann, J.} and {Conselice, C. J.} and {Dole, H.} and {Cabanac, R.} and {Courtois, H. M.} and {Castander, F. J.} and {Duc, P. A.} and {Fosalba, P.} and {Guinet, D.} and {Kruk, S.} and {Kuchner, U.} and {Serrano, S.} and {Soubrie, E.} and {Tramacere, A.} and {Wang, L.} and {Amara, A.} and {Auricchio, N.} and {Bender, R.} and {Bodendorf, C.} and {Bonino, D.} and {Branchini, E.} and {Brau-Nogue, S.} and {Brescia, M.} and {Capobianco, V.} and {Carbone, C.} and {Carretero, J.} and {Cavuoti, S.} and {Cimatti, A.} and {Cledassou, R.} and {Congedo, G.} and {Conversi, L.} and {Copin, Y.} and {Corcione, L.} and {Costille, A.} and {Cropper, M.} and {Da Silva, A.} and {Degaudenzi, H.} and {Douspis, M.} and {Dubath, F.} and {Duncan, C. A. J.} and {Dupac, X.} and {Dusini, S.} and {Farrens, S.} and {Ferriol, S.} and {Frailis, M.} and {Franceschi, E.} and {Fumana, M.} and {Garilli, B.} and {Gillard, W.} and {Gillis, B.} and {Giocoli, C.} and {Grazian, A.} and {Grupp, F.} and {Haugan, S. V. H.} and {Holmes, W.} and {Hormuth, F.} and {Hudelot, P.} and {Jahnke, K.} and {Kermiche, S.} and {Kiessling, A.} and {Kilbinger, M.} and {Kitching, T.} and {Kohley, R.} and {Kümmel, M.} and {Kunz, M.} and {Kurki-Suonio, H.} and {Ligori, S.} and {Lilje, P. B.} and {Lloro, I.} and {Maiorano, E.} and {Mansutti, O.} and {Marggraf, O.} and {Markovic, K.} and {Marulli, F.} and {Massey, R.} and {Maurogordato, S.} and {Melchior, M.} and {Meneghetti, M.} and {Meylan, G.} and {Moresco, M.} and {Morin, B.} and {Moscardini, L.} and {Munari, E.} and {Nakajima, R.} and {Niemi, S. M.} and {Padilla, C.} and {Paltani, S.} and {Pasian, F.} and {Pedersen, K.} and {Pettorino, V.} and {Pires, S.} and {Poncet, M.} and {Popa, L.} and {Pozzetti, L.} and {Raison, F.} and {Rebolo, R.} and {Rhodes, J.} and {Roncarelli, M.} and {Rossetti, E.} and {Saglia, R.} and {Schneider, P.} and {Secroun, A.} and {Seidel, G.} and {Sirignano, C.} and {Sirri, G.} and {Stanco, L.} and {Starck, J.-L.} and {Tallada-Crespí, P.} and {Taylor, A. N.} and {Tereno, I.} and {Toledo-Moreo, R.} and {Torradeflot, F.} and {Valentijn, E. A.} and {Valenziano, L.} and {Wang, Y.} and {Welikala, N.} and {Weller, J.} and {Zamorani, G.} and {Zoubian, J.} and {Baldi, M.} and {Bardelli, S.} and {Camera, S.} and {Farinelli, R.} and {Medinaceli, E.} and {Mei, S.} and {Polenta, G.} and {Romelli, E.} and {Tenti, M.} and {Vassallo, T.} and {Zacchei, A.} and {Zucca, E.} and {Baccigalupi, C.} and {Balaguera-Antolínez, A.} and {Biviano, A.} and {Borgani, S.} and {Bozzo, E.} and {Burigana, C.} and {Cappi, A.} and {Carvalho, C. S.} and {Casas, S.} and {Castignani, G.} and {Colodro-Conde, C.} and {Coupon, J.} and {de la Torre, S.} and {Fabricius, M.} and {Farina, M.} and {Ferreira, P. G.} and {Flose-Reimberg, P.} and {Fotopoulou, S.} and {Galeotta, S.} and {Ganga, K.} and {Garcia-Bellido, J.} and {Gaztanaga, E.} and {Gozaliasl, G.} and {Hook, I. M.} and {Joachimi, B.} and {Kansal, V.} and {Kashlinsky, A.} and {Keihanen, E.} and {Kirkpatrick, C. C.} and {Lindholm, V.} and {Mainetti, G.} and {Maino, D.} and {Maoli, R.} and {Martinelli, M.} and {Martinet, N.} and {McCracken, H. J.} and {Metcalf, R. B.} and {Morgante, G.} and {Morisset, N.} and {Nightingale, J.} and {Nucita, A.} and {Patrizii, L.} and {Potter, D.} and {Renzi, A.} and {Riccio, G.} and {Sánchez, A. G.} and {Sapone, D.} and {Schirmer, M.} and {Schultheis, M.} and {Scottez, V.} and {Sefusatti, E.} and {Teyssier, R.} and {Tutusaus, I.} and {Valiviita, J.} and {Viel, M.} and {Whittaker, L.} and {Knapen, J. H.}},
	title = {Euclid preparation - XIII. Forecasts for galaxy morphology with the Euclid Survey using deep generative models},
	DOI= "10.1051/0004-6361/202141393",
	url= "https://doi.org/10.1051/0004-6361/202141393",
	journal = {A\&A},
	year = 2022,
	volume = 657,
	pages = "A90",
}

@article{Peng_2010,
doi = {10.1088/0004-6256/139/6/2097},
url = {https://dx.doi.org/10.1088/0004-6256/139/6/2097},
year = {2010},
month = {apr},
publisher = {The American Astronomical Society},
volume = {139},
number = {6},
pages = {2097},
author = {Chien Y. Peng and Luis C. Ho and Chris D. Impey and Hans-Walter Rix},
title = {DETAILED DECOMPOSITION OF GALAXY IMAGES. II. BEYOND AXISYMMETRIC MODELS},
journal = {The Astronomical Journal}
}

@ARTICLE{2015AandC....10..121R,
       author = {{Rowe}, B.~T.~P. and {Jarvis}, M. and {Mandelbaum}, R. and {Bernstein}, G.~M. and {Bosch}, J. and {Simet}, M. and {Meyers}, J.~E. and {Kacprzak}, T. and {Nakajima}, R. and {Zuntz}, J. and {Miyatake}, H. and {Dietrich}, J.~P. and {Armstrong}, R. and {Melchior}, P. and {Gill}, M.~S.~S.},
        title = "{GALSIM: The modular galaxy image simulation toolkit}",
      journal = {Astronomy and Computing},
         year = 2015,
        month = {apr},
       volume = {10},
        pages = {121-150},
          doi = {10.1016/j.ascom.2015.02.002},
archivePrefix = {arXiv},
       eprint = {1407.7676},
 primaryClass = {astro-ph.IM},
       adsurl = {https://ui.adsabs.harvard.edu/abs/2015A&C....10..121R}
}

@article{CHEN2024100825,
title = {Galmoss: A package for GPU-accelerated galaxy profile fitting},
journal = {Astronomy and Computing},
volume = {47},
pages = {100825},
year = {2024},
issn = {2213-1337},
doi = {https://doi.org/10.1016/j.ascom.2024.100825},
url = {https://www.sciencedirect.com/science/article/pii/S2213133724000404},
author = {Mi Chen and Rafael {S. de Souza} and Quanfeng Xu and Shiyin Shen and Ana L. Chies-Santos and Renhao Ye and Marco A. Canossa-Gosteinski and Yanping Cong}
}

@article{10.1111/j.1365-2966.2008.13936.x,
    author = {Joshi, Himali and Kumar, Brijesh and Singh, K. P. and Sagar, Ram and Sharma, Saurabh and Pandey, J. C.},
    title = {Multiwavelength study of a young open cluster NGC 7419},
    journal = {Monthly Notices of the Royal Astronomical Society},
    volume = {391},
    number = {3},
    pages = {1279-1300},
    year = {2008},
    month = {11},
    issn = {0035-8711},
    doi = {10.1111/j.1365-2966.2008.13936.x},
    url = {https://doi.org/10.1111/j.1365-2966.2008.13936.x},
    eprint = {https://academic.oup.com/mnras/article-pdf/391/3/1279/17320341/mnras391-1279.pdf},
}

@article{Wuyts_2011,
doi = {10.1088/0004-637X/742/2/96},
url = {https://dx.doi.org/10.1088/0004-637X/742/2/96},
year = {2011},
month = {nov},
publisher = {The American Astronomical Society},
volume = {742},
number = {2},
pages = {96},
author = {Wuyts, Stijn and Förster Schreiber, Natascha M. and van der Wel, Arjen and Magnelli, Benjamin and Guo, Yicheng and Genzel, Reinhard and Lutz, Dieter and Aussel, Hervé and Barro, Guillermo and Berta, Stefano and Cava, Antonio and Graciá-Carpio, Javier and Hathi, Nimish P. and Huang, Kuang-Han and Kocevski, Dale D. and Koekemoer, Anton M. and Lee, Kyoung-Soo and Le Floc'h, Emeric and McGrath, Elizabeth J. and Nordon, Raanan and Popesso, Paola and Pozzi, Francesca and Riguccini, Laurie and Rodighiero, Giulia and Saintonge, Amelie and Tacconi, Linda},
title = {GALAXY STRUCTURE AND MODE OF STAR FORMATION IN THE SFR–MASS PLANE FROM z ~ 2.5 TO z ~ 0.1},
journal = {The Astrophysical Journal}
}

@ARTICLE{Ahn2012,
       author = {{Ahn}, Christopher P. and {Alexandroff}, Rachael and {Allende Prieto}, Carlos and {Anderson}, Scott F. and {Anderton}, Timothy and {Andrews}, Brett H. and {Aubourg}, {\'E}ric and {Bailey}, Stephen and {Balbinot}, Eduardo and {Barnes}, Rory and {Bautista}, Julian and {Beers}, Timothy C. and {Beifiori}, Alessandra and {Berlind}, Andreas A. and {Bhardwaj}, Vaishali and {Bizyaev}, Dmitry and {Blake}, Cullen H. and {Blanton}, Michael R. and {Blomqvist}, Michael and {Bochanski}, John J. and {Bolton}, Adam S. and {Borde}, Arnaud and {Bovy}, Jo and {Brandt}, W.~N. and {Brinkmann}, J. and {Brown}, Peter J. and {Brownstein}, Joel R. and {Bundy}, Kevin and {Busca}, N.~G. and {Carithers}, William and {Carnero}, Aurelio R. and {Carr}, Michael A. and {Casetti-Dinescu}, Dana I. and {Chen}, Yanmei and {Chiappini}, Cristina and {Comparat}, Johan and {Connolly}, Natalia and {Crepp}, Justin R. and {Cristiani}, Stefano and {Croft}, Rupert A.~C. and {Cuesta}, Antonio J. and {da Costa}, Luiz N. and {Davenport}, James R.~A. and {Dawson}, Kyle S. and {de Putter}, Roland and {De Lee}, Nathan and {Delubac}, Timoth{\'e}e and {Dhital}, Saurav and {Ealet}, Anne and {Ebelke}, Garrett L. and {Edmondson}, Edward M. and {Eisenstein}, Daniel J. and {Escoffier}, S. and {Esposito}, Massimiliano and {Evans}, Michael L. and {Fan}, Xiaohui and {Femen{\'\i}a Castell{\'a}}, Bruno and {Fern{\'a}ndez Alvar}, Emma and {Ferreira}, Leticia D. and {Filiz Ak}, N. and {Finley}, Hayley and {Fleming}, Scott W. and {Font-Ribera}, Andreu and {Frinchaboy}, Peter M. and {Garc{\'\i}a-Hern{\'a}ndez}, D.~A. and {Garc{\'\i}a P{\'e}rez}, A.~E. and {Ge}, Jian and {G{\'e}nova-Santos}, R. and {Gillespie}, Bruce A. and {Girardi}, L{\'e}o and {Gonz{\'a}lez Hern{\'a}ndez}, Jonay I. and {Grebel}, Eva K. and {Gunn}, James E. and {Guo}, Hong and {Haggard}, Daryl and {Hamilton}, Jean-Christophe and {Harris}, David W. and {Hawley}, Suzanne L. and {Hearty}, Frederick R. and {Ho}, Shirley and {Hogg}, David W. and {Holtzman}, Jon A. and {Honscheid}, Klaus and {Huehnerhoff}, J. and {Ivans}, Inese I. and {Ivezi{\'c}}, {\v{Z}}eljko and {Jacobson}, Heather R. and {Jiang}, Linhua and {Johansson}, Jonas and {Johnson}, Jennifer A. and {Kauffmann}, Guinevere and {Kirkby}, David and {Kirkpatrick}, Jessica A. and {Klaene}, Mark A. and {Knapp}, Gillian R. and {Kneib}, Jean-Paul and {Le Goff}, Jean-Marc and {Leauthaud}, Alexie and {Lee}, Khee-Gan and {Lee}, Young Sun and {Long}, Daniel C. and {Loomis}, Craig P. and {Lucatello}, Sara and {Lundgren}, Britt and {Lupton}, Robert H. and {Ma}, Bo and {Ma}, Zhibo and {MacDonald}, Nicholas and {Mack}, Claude E. and {Mahadevan}, Suvrath and {Maia}, Marcio A.~G. and {Majewski}, Steven R. and {Makler}, Martin and {Malanushenko}, Elena and {Malanushenko}, Viktor and {Manchado}, A. and {Mandelbaum}, Rachel and {Manera}, Marc and {Maraston}, Claudia and {Margala}, Daniel and {Martell}, Sarah L. and {McBride}, Cameron K. and {McGreer}, Ian D. and {McMahon}, Richard G. and {M{\'e}nard}, Brice and {Meszaros}, Sz. and {Miralda-Escud{\'e}}, Jordi and {Montero-Dorta}, Antonio D. and {Montesano}, Francesco and {Morrison}, Heather L. and {Muna}, Demitri and {Munn}, Jeffrey A. and {Murayama}, Hitoshi and {Myers}, Adam D. and {Neto}, A.~F. and {Nguyen}, Duy Cuong and {Nichol}, Robert C. and {Nidever}, David L. and {Noterdaeme}, Pasquier and {Nuza}, Sebasti{\'a}n E. and {Ogando}, Ricardo L.~C. and {Olmstead}, Matthew D. and {Oravetz}, Daniel J. and {Owen}, Russell and {Padmanabhan}, Nikhil and {Palanque-Delabrouille}, Nathalie and {Pan}, Kaike and {Parejko}, John K. and {Parihar}, Prachi and {P{\^a}ris}, Isabelle and {Pattarakijwanich}, Petchara and {Pepper}, Joshua and {Percival}, Will J. and {P{\'e}rez-Fournon}, Ismael and {P{\'e}rez-R{\`a}fols}, Ignasi and {Petitjean}, Patrick and {Pforr}, Janine and {Pieri}, Matthew M. and {Pinsonneault}, Marc H. and {Porto de Mello}, G.~F. and {Prada}, Francisco and {Price-Whelan}, Adrian M. and {Raddick}, M. Jordan and {Rebolo}, Rafael and {Rich}, James and {Richards}, Gordon T. and {Robin}, Annie C. and {Rocha-Pinto}, Helio J. and {Rockosi}, Constance M. and {Roe}, Natalie A. and {Ross}, Ashley J. and {Ross}, Nicholas P. and {Rossi}, Graziano and {Rubi{\~n}o-Martin}, J.~A. and {Samushia}, Lado and {Sanchez Almeida}, J. and {S{\'a}nchez}, Ariel G. and {Santiago}, Bas{\'\i}lio and {Sayres}, Conor and {Schlegel}, David J. and {Schlesinger}, Katharine J. and {Schmidt}, Sarah J. and {Schneider}, Donald P. and {Schultheis}, Mathias and {Schwope}, Axel D. and {Sc{\'o}ccola}, C.~G. and {Seljak}, Uros and {Sheldon}, Erin and {Shen}, Yue and {Shu}, Yiping and {Simmerer}, Jennifer and {Simmons}, Audrey E. and {Skibba}, Ramin A. and {Skrutskie}, M.~F. and {Slosar}, A. and {Sobreira}, Flavia and {Sobeck}, Jennifer S. and {Stassun}, Keivan G. and {Steele}, Oliver and {Steinmetz}, Matthias and {Strauss}, Michael A. and {Streblyanska}, Alina and {Suzuki}, Nao and {Swanson}, Molly E.~C. and {Tal}, Tomer and {Thakar}, Aniruddha R. and {Thomas}, Daniel and {Thompson}, Benjamin A. and {Tinker}, Jeremy L. and {Tojeiro}, Rita and {Tremonti}, Christy A. and {Vargas Maga{\~n}a}, M. and {Verde}, Licia and {Viel}, Matteo and {Vikas}, Shailendra K. and {Vogt}, Nicole P. and {Wake}, David A. and {Wang}, Ji and {Weaver}, Benjamin A. and {Weinberg}, David H. and {Weiner}, Benjamin J. and {West}, Andrew A. and {White}, Martin and {Wilson}, John C. and {Wisniewski}, John P. and {Wood-Vasey}, W.~M. and {Yanny}, Brian and {Y{\`e}che}, Christophe and {York}, Donald G. and {Zamora}, O. and {Zasowski}, Gail and {Zehavi}, Idit and {Zhao}, Gong-Bo and {Zheng}, Zheng and {Zhu}, Guangtun and {Zinn}, Joel C.},
        title = "{The Ninth Data Release of the Sloan Digital Sky Survey: First Spectroscopic Data from the SDSS-III Baryon Oscillation Spectroscopic Survey}",
      journal = {\apjs},
         year = 2012,
        month = {dec},
       volume = {203},
       number = {2},
          eid = {21},
        pages = {21},
          doi = {10.1088/0067-0049/203/2/21},
archivePrefix = {arXiv},
       eprint = {1207.7137},
 primaryClass = {astro-ph.IM},
       adsurl = {https://ui.adsabs.harvard.edu/abs/2012ApJS..203...21A}
}

@ARTICLE{1926ApJ....64..321H,
       author = {{Hubble}, E.~P.},
        title = "{Extragalactic nebulae.}",
      journal = {\apj},
         year = 1926,
        month = 12,
       volume = {64},
        pages = {321-369},
          doi = {10.1086/143018}
}

@article{Conselice2014,
  title={The evolution of galaxy structure over cosmic time},
  author={Conselice, Christopher J},
  journal={Annual Review of Astronomy and Astrophysics},
  volume={52},
  number={1},
  pages={291--337},
  year={2014},
  publisher={Annual Reviews}
}

@article{Wang2004,
  title={Image quality assessment: from error visibility to structural similarity},
  author={Wang, Zhou and Bovik, Alan C and Sheikh, Hamid R and Simoncelli, Eero P},
  journal={IEEE transactions on image processing},
  volume={13},
  number={4},
  pages={600--612},
  year={2004},
  publisher={IEEE}
}

\PublishersNote{}
\end{adjustwidth}
\end{document}